\documentclass[aps,pre,preprint,showpacs]{revtex4}
\usepackage{epsf}
\usepackage{amsmath,amssymb}
\usepackage{graphicx}
\usepackage{color}
\usepackage[T2A]{fontenc}
\usepackage[russian,english]{babel}

\begin{document}

\vspace{0.5cm}
\title{Chaotic Interference and Quantum-Classical Correspondence: Mechanisms of
Decoherence and State Mixing}
\author{Valentin V. Sokolov}
\affiliation{Budker Institute of Nuclear Physics, Novosibirsk, Russia}
\affiliation{Novosibirsk State Technical University, Novosibirsk, Russia}
\author{Oleg V. Zhirov}
\affiliation{Budker Institute of Nuclear Physics, Novosibirsk, Russia}
\affiliation{Novosibirsk State University, Novosibirsk, Russia}

\date{\today}



\maketitle

\section{Outline}
\label{sec:outline}

\noindent
\looseness-1

    The famous Nils Bohr's quantum-classical correspondence
principle states that the classical mechanics is a limiting case of the
more general quantum mechanics. This implies that ``under certain
conditions" quantum laws of motion become equivalent to classical
ones. One of the conditions is fairly obvious: the corresponding
classical action should be very large as compared with the Planck's
constant $\hbar$. But is this the \textit{sufficient} condition? In fact,
\textit{it is not!}

The quantum laws show up in two different, although not entirely independent ways:
\begin{enumerate}
    \item  discrete spectrum of finite motion,
    \item  interference phenomena.
 \end{enumerate}
Even if the energy spectrum of a finite closed quantum system
becomes continuous in the formal limit of vanishing $\hbar$, the
interference effects cannot disappear in similar manner. Indeed, a
quantum wave functions has no definite classical counterpart.
Meanwhile, suppression of effects of quantum interference
("decoherence") is a key requirement for the classical laws to appear.
Being, in essence, of quite general nature, this problem takes on
special significance in the non-trivial case of non-linear classically
chaotic quantum systems.

A number of typical manifestations of the quantum coherence in the
time evolution as well as eigenstates' properties are widely discussed
in the scientific literature:
\begin{itemize}
\item Wave packet dynamics and decay of quantum fidelity.
\item Universal local spectral fluctuations.
\item Scars in the stationary eigenfunctions.
\item Elastic enhancement in chaotic resonance scattering.
\item Weak localization in transport phenomena.
\end{itemize}

The specific features of quantum dynamics of classically chaotic
systems seem to be  in striking contrast with those of
genuine classical chaos. Since these features can even question the
validity of the quantum-classical correspondence principle by itself,
a more profound analysis  is needed for  understanding
the bridge between the classical and quantum chaotic worlds.

\subsection{Chaotic time evolution}
\label{sec:ChaoTimeEvolution}
    In the case of regular classical dynamics, the system's response to a
weak external perturbation is proportional to its strength and the
system may be still treated as a \textit{closed} one during
sufficiently long time. In contrast, chaotic classical dynamics is
exponentially unstable and therefore it is extremely sensitive to any
uncontrollable external influence. We can never neglect
the influence of the environment. This therefore stipulates the
\textit{self-mixing} property of classical dynamical chaos and, as a
consequence, a very fast decay of the phase correlations (here and
throughout the paper we use the language of action-angle variables).

Whereas the exponential decay of the phase correlations is an
underlying feature of the classical dynamical chaos \cite{Chiri79},
the so called "quantum chaos", i.e. quantum dynamics
of classically chaotic systems is not by itself capable of destroying
the \textit{quantum phase} coherence. Strictly speaking, any
initially pure quantum state remains pure during arbitrary long
evolution. Quantization of the phase space removes exponential
instability and makes the quantum dynamics to be substantially more
stable than the classical one. There exists a threshold $\sigma_c(t)$
of the external noise strength $\sigma$ below which the coherence
survives   up to the time $t$ \cite{Sokol08,Sokol08a,Sokol09}
Only appreciably strong noise or finite measurement accuracy can
produce a mixture of quantum states sufficient for noticeable
suppression of quantum coherence. A number of appropriate
characteristics: Peres fidelity $F(t)$, purity $\mathbb{P}(t)$,
Shannon (information) ${\cal I}(t)$ and von Neumann (correlational)
${\cal S}(t)$ entropies are used to demonstrate the gradual loss of
quantum coherence during system's evolution in the presence of
noisy background. Being sensitive to the quantum coherence, the
von Neumann entropy remains smaller than the Shannon entropy but
runs up monotonically to the latter when the evolution time
approaches some moment $t_{(dec)}$ when decoherence becomes
complete. After this time, the system occupies the whole phase space
volume accessible at the running degree of excitation energy thus
reaching a sort of equilibrium. Henceforth the phase volume expands
``adiabatically'': the both entropies remain almost constant. The
evolution after the time
$t_{(dec)}$ is Markovian \cite{Sokol09}.

\subsection{2D billiards}
\label{sec:2Dbilliards}
    The plain ($2D$) areas with closed irregular borders called billiards
are pet systems often used to illustrate characteristic features of the
classical dynamical chaos. With the advent of the ability to fabricate
mesoscopic analogs of the classical billiards the opportunity appeared
to observe experimentally the signatures of chaos in classically
chaotic quantum systems. An excellent possibility thus has arisen to
verify the theoretical concepts developed in numerous theoretical
investigations of "quantum chaos" phenomena. Extensive study of the
electron transport through ballistic $2D$ meso-structures
\cite{Jalab90,Marcu92} (see also \cite{Beena97,Alhas00} and
references therein) have fully confirmed correctness of the basic
ideas \cite{Izrai90,Gutzw90,Haake91} of the theory of the chaotic
quantum interference and relevance \cite{Baran94,Beena97,Alhas00}
of the so called random-matrix approach
\cite{Verba85,Mello85,Mehta91} to the problem of the universal
spectral fluctuations as well as conductance fluctuations in open
mesoscopic set-ups (see sec.\ref{sec:Billiards&Transport}).

The energy $E$ is the only integral of motion in a closed billiard.
Repeating reflections from the border produces a countless manifold
of unclosed exponentially unstable trajectories and, as a consequence,
chaotic dynamics. At the same time, there exists a countable set of
specific, but also exponentially unstable, periodic orbits. The number
$N(T)$ of the latter trajectories grows exponentially with the period
$T$ as $N(T)\sim e^{\frac{T}{\tau_c}}$ with the characteristic time
$\tau_c$ directly connected, $\tau_c\thicksim 1/\Lambda$,  to the
Lyapunov exponent $\Lambda$ that describes the exponential
instability \cite{Zasla85}. Meanwhile, in contravention with the
classical exponential instability, stable interference maxima (scars)
along the periodic classical orbits (alongside with irregularly
scattered point-like interference maxima (specks)) are discovered
numerically \cite{Helle84} in the stationary billiard's eigenstates
even in the very deep semi-classical region . Does this fact
compromises the quantum-classical correspondence principle?\\

To answer afore-posed question let us consider first an elementary
example of $1D$ finite motion. A particle with a mass $m$ and
energy $E$ moves in a potential well. The classical position
probability density is proportional to the fraction of the period of
oscillations the particle spends near a point $x$ and is easily found to
be:
\begin{equation}
  w_c(x)=
  \left\{
    \begin{array}{ll}
        0 &x\notin [a, b],\\
        \frac{m\,\omega_c(E)}{\pi\, p_c(x)},\,\,\, &x\in[a, b]\,.\\
    \end{array}
  \right.
  \label{eq:classW}
\end{equation}
where $p_c(x)=\sqrt{2m\left[E-U(x)\right]}$ is the particle's classical
momentum at the point $x$ and $\omega_c(E)$ is the frequency of
classical oscillations between the turning points $a$ and $b$.

On the other hand the semi-classical solution of the corresponding
quantum problem yields
\begin{eqnarray}
  w_n(x)&=&|\psi_n(x)|^2 \nonumber \\
    &=&\left\{
        \begin{array}{ll}
          e^{\{-\frac{2}{\hbar}\int_x^a dx' |p_c(x')|\}},
            e^{\{-\frac{2}{\hbar}\int_b^x dx' |p_c(x')|\}} &\rightarrow 0,\\
          \frac{m\,\omega_c(E_n)}{\pi\, p_c(x)}\,2\cos^2\{\frac{1}{\hbar}
            \int_a^x dx'p_c(x')-\frac{\pi}{4}\} &\rightarrow ?
        \end{array}
      \right.
   \label{eq:quantW}
\end{eqnarray}
\begin{figure}[h]
\begin{center}
\includegraphics[width=70mm,angle=0,
   keepaspectratio=true]{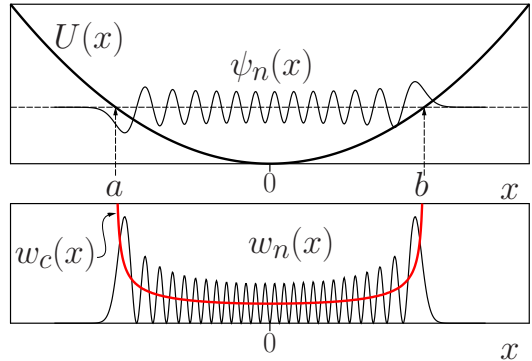}
\caption{\textit{Top}: wave function of harmonic oscillator for the
energy level $n=25$; \textit{bottom}: corresponding probability
distributions are shown by thick red and thin black lines for classical and quantum cases respectively.}
\label{fig:1DWell}
\end{center}
\end{figure}
As expected (see the simplest of linear oscillator,
Fig.\ref{fig:1DWell}, bottom frame),
the probability density (\ref{eq:quantW}) vanishes in
the classically forbidden region when the semi-classical parameter
goes to infinity but there exists no reasonable result in the
\textit{classically allowed} interval. The wildly fluctuating without
approaching a certain limit cosine-square factor appears from the
interference of two waves running in the opposite directions. To
attain the classical result (\ref{eq:classW})  an
additional \textit{averaging} over some finite either
position  $\Delta x$ or energy $\Delta n$ intervals around fixed
 $|x\rangle$ or $|n\rangle$ states is necessary.
\begin{eqnarray}
\label{eq:avWn_x}
  \overline{w_n(x)}^{(x)}&=&\langle n|\left(\int_{x'\in\Delta x} d x'\,
    \mathbf{p}_x(x')|x'\rangle\langle x'|\right)|n\rangle=
    \langle n|{\hat\rho}^{(x)}|n\rangle\\
  \label{eq:avWn_n}
  \overline{w_n(x)}^{(n)}&=&
     \langle x|\left(\sum_{n'\in \Delta n}\mathbf{p}_{n}(n')
      |n'\rangle\langle n'|\right)|x\rangle=
     \langle x|{\hat\rho}^{(n)}|x\rangle\,.
\end{eqnarray}
The density matrices ${\hat\rho}^{(q)}$ $(q=n,x)$ describe
{\it  incoherent mixtures} of the quantum states within the indicated
intervals. The real and normalized to unity quantities $\mathbf{p}$
characterize the weights of these states. Obviously, the range of
averaging $\Delta x$ should satisfy the condition
$\Delta x\geqslant \pi\hbar/p_c(x)$ to meet the classical behavior.
Similar reasoning leads to the condition
$\Delta E\approx \Delta n\,\hbar \omega_c\geqslant 2\pi\hbar/T$
where $T=2\int_a^b d x/v_c(x)$ is the period of classical
oscillations. Similarly, averaging over the energy interval $\Delta E$
wipes off all scars of eigenstates of a $2D$ billiard with the periods
$T>\hbar/\Delta E$, whereas those with smaller periods still survive
\cite{Bogom88}.

Strictly speaking, decoherence is not perfect as long as off-diagonal
matrix elements of the density matrix still exist. They are complex
and their phases carry some information on the more subtle
interference effects. Complete decoherence is achieved only when the
number of  mixing states is so large that the density matrix becomes
proportional to the unit matrix and ceases to depend on the basis in
the Hilbert space of states. As a matter of fact, decoherence can
originate only from: (i) the process of preparation of initially mixed
state, and (ii) mixing induced during the time evolution by a
persisting external noise.

\subsection{The basics of quantum mixed states.}
\label{sec:basicsQuMixStates}
    The concept of mixed states plays a paramount role in the problem of
decoherence. A pure quantum state is specified at some moment of
time $t$ by its wave vector $|\psi(t)\rangle$ in the Hilbert space. This
allows, in particular, calculation of the mean value
$O(t)=\langle\psi(t)|{\hat O}|\psi(t)\rangle$
(in what follows we suggest this vector to be normalized to unity,
$\langle\psi(t)|\psi(t)\rangle$=1) of any dynamical quantity
represented by a Hermitian operator ${\hat O}$. Equivalently, the
same state can be described by the density matrix
${\hat\rho (t)=|\psi(t)\rangle}\langle\psi(t)|$
that satisfies two obvious conditions:
\begin{equation}\label{P_rho}
  Tr{\hat\rho }(t)=1,\quad {\hat\rho}^2(t)={\hat\rho}(t).
\end{equation}
The second relation is the necessary and sufficient condition of purity
of a quantum state. With these definitions at hands, the mean values
can be presented in one of the following forms
\begin{equation}\label{mean_O}
O(t)=\langle\psi(t)|{\hat O}|\psi(t)\rangle=
         Tr\left[{\hat O}\,{\hat\rho} (t)\right]=
         \sum_{m,n}O_{n m}(t)\rho_{m n}(t).
\end{equation}
The expression in the right hand side represents the same mean value
in the basis of a complete set of motionless vectors $|n\rangle$ in the
Hilbert space. Since the density matrix is Hermitian, all its diagonal
elements $\rho_{n n}(t)$ are real.

Like any Hermitian matrix, the density matrix $\rho_{m n}(t)$ can be
diagonalized with the help of some unitary transformation. On
account of the conditions (\ref{P_rho}) the density matrix of a pure
quantum state has only one nonzero eigenvalue that equals one.
Obviously, the diagonalization returns us to the original form
${\hat\rho(t)= |\psi(t)\rangle}\langle\psi(t)|$. The transformation
matrix depends on time if the system undergo time evolution. In
practice, a fixed basis of states $|n\rangle$ is, as a rule, a more
relevant choice from physical point of view.

A mixed quantum state is described at any moment of time $t$ by an
incoherent sum
\begin{equation}\label{mixed_rho}
  {\hat\rho}(t)=\sum_k \mathbf{p}_k(t)|v_k(t)\rangle\langle v_k(t)|
\end{equation}
of binary contributions where the states $|v_k(t)\rangle$ are the
eigenvectors of the density matrix when $0\leq\mathbf{p}_k(t)\leq 1$
stand for the weights of the corresponding pure fragments. These
weights do not actually depend on time during dynamical evolution
described by some unitary evolution matrix ${\hat U}(t)$. It is not the
case, however, if the system interacts with a noisy background.

In the fixed basis $|n\rangle$, an initially
diagonal incoherent mixture develops off-diagonal elements even
during unitary dynamical evolution. There exists, however, a
convenient invariant measure of state purity called \textit{Purity}
\begin{equation}\label{Purity}
 \mathbb{P}(t)=\mathrm{Tr}\left[{\hat\rho}^2(t)\right]=
                           \sum_k\mathbf{p}^2_k= \mathbb{P}(0)
\end{equation}
that remains constant as long as the noise is absent. The Purity is
restricted to the interval $0\leq\mathbb{P}\leq 1$ and is mounting to
one in the limit of a perfectly pure state. In many respects similar
properties are inherent in the invariant von Neumann entropy
\begin{equation}\label{v_N_ent}
  {\cal S}(t)=-\mathrm{Tr}\left[{\hat\rho}(t)\ln{\hat\rho}(t)\right]=
                      {\cal S}(0)
\end{equation}
which vanishes only in the case of a perfectly pure state.

It is very useful to transfer the story on the, generally mixed,
quantum states evolution to the language of the phase space
\cite{Agarw70,Agarw70a,Sokol08a}. This elucidates analogy  and
distinctions between classical and quantum dynamics. A double
Fourier transformation of the density matrix
\begin{equation}\label{Wfunc}
\begin{array}{c}
W(\alpha^*,\alpha;t)=
\frac{1}{\pi^2\hbar}\int d^2\eta\,
\exp\left(\eta\frac{\alpha^\star}{\sqrt\hbar}-
\eta^\star\frac{\alpha}{\sqrt\hbar}\right)
\tilde{\rho}(\eta^\star,\eta;t)=\\
\frac{1}{\pi^2\hbar}\int d^2\eta\,
\exp\left(\eta^*\frac{\alpha}{\sqrt\hbar}-
\eta\frac{\alpha^*}{\sqrt\hbar}\right) \mathrm{Tr}\left[{\hat\rho(t)}\,
{\hat D(\eta)}\right]\,,
\end{array}
\end{equation}
with the operator
${\hat D(\eta)}=\exp(\eta\,{\hat a}^{\dag}-\eta^*{\hat a})$
being the coherent states displacement
defines the {\it Wigner function} $W(\alpha^*,\alpha;t)$
that is a direct quantum counterpart of the classical phase space
distribution function $W^{(c)}(\alpha^*,\alpha;t)$. The complex
variables $\alpha^*,\alpha$ are connected with the standard
action-angle variables $I, \theta$ by the canonical transformations
$\alpha=\sqrt{I}e^{-i\theta},\,\alpha^*=\sqrt{I}e^{i\theta}$.

Deep and detailed analysis of properties of the density matrix and its
significant role for the problem of decoherence has been presented in
the review \cite{Zurek03}.

\subsection{Peres fidelity}
\label{sec:PeresFidelity}
    Response of an evolving in time quantum system to a weak external
perturbation is of prime interest in the context of the problem of
stability and reversibility of quantum motion. The customary
quantitative characteristic of sensitivity of classically chaotic
quantum dynamics to such perturbations is the Peres fidelity
\cite{Peres84}:
\begin{equation}\label{Fidelity}
  F(t)=\frac{\mathrm{Tr}\left[{\hat\rho}_H(t)\,
           {\hat\rho}_{H'}(t)\right]}{\mathbb{P}(t)}=
            \frac{\sum_{k l}{\bf p}_k{\bf p}'_l\big|\langle
            v_k(t)|v'_l(t)\rangle\big|^2}{\mathbb{P}(t)}\,.
\end{equation}
Fidelity measures the weighted mean distance between two,
generally mixed, quantum states evolving according to slightly
different Hamiltonians ${\hat H}$ and ${\hat H}'$, is bounded in the
interval $[0, 1]$ and equals one when ${\hat H}'={\hat H}$. Its time
decay due to diminishing of the overlap of $v$-eigenvectors
elucidates the sensitivity of the motion to an external influence. In
particular, this quantity enable one to directly connect stability and
reversibility of quantum dynamics with complexity of the quantum
Wigner function. More than that, the notion of the Peres fidelity
directly extends to the classical mechanics
(see sec.II  in \cite{Benen03} and \cite{Benen03a}).
The number of $\theta$-harmonics can serve (see below) as a natural
quantitative measure of complexity of the both, classical as well as
quantum phase space (quasi-)distributions
\cite{Sokol08a} (see, however, \cite{Benen12} where another measure of complexity has been proposed that is relevant in the case of systems with more than one degrees of freedom).

\subsection{Open mesoscopic billiards and electron quantum transport}
\label{sec:Billiards&Transport}
    Scattering of quantum particles by 2D billiard-like mesoscopic
structures connected to the continuum by two opening leads each supporting
$M$ channels is an intriguing issue that attracted a lot of attention of
theorists as well as experimentalists for at least the last two decades.
If the mean free path of such a particle exceeds the typical size of the
structure the particle's dynamics inside it strongly depends on the shape
of its border. The classical motion in this case becomes   stochastic and
quantum scattering is expected to be described well within the
framework of the random matrix approach to the resonance
scattering theory. Intensive experimental studies confirmed in many
respects these expectations. Nevertheless, electron transport
experiments with ballistic quantum dots reveal noticeable and
persisting up to zero temperature loss of the quantum-mechanical
coherence in contravention of predictions of the random-matrix as
well as semiclassical scattering theories.

There exists a number of different methods of accounting for the
decoherence effect in the ballistic quantum transport processes.As
a matter of fact, all of them originate from the pioneering
B\"uttiker's ideas \cite{Buetti86}. The dot is supposed to be
connected someway with a bath of electrons so that the
dot and the bath can exchange electrons in such a manner that the
mean exchange electric current vanishes. Since the incoming electron
carries no information of the phase of the wave function in the dot,
the coherence turns our to be suppressed. The cost payed is that the
number of particle is conserved only in average during the scattering
so that the scattering matrix is not, strictly speaking, unitary.
An alternative approach has been therefore proposed in
\cite{Beena05} with a closed long stub instead of an opening lead.
Thereby unitarity of the $S$-matrix is guaranteed and none of the
electrons is lost at any individual act of scattering.

Decoherence
takes place in this case because of a spatially random time-dependent
external electric field that acts in the stub. As a result, an electron
once penetrated the stub return back in the dot without whatever phase
memory.

In spite of the advantages of the stub model the necessity of
introducing {\it ad hoc} an external time dependent potential seems
to be somewhat artificial. Still another possibility arises if a
time-independent weak interaction is taken into account with a
relatively rare irregular impurities in the semiconductor
heterostructure to whose interface region the electrons are confined.
At that, unavoidable experimental averaging over the scattering energy
on account of finite experimental accuracy in measuring the cross sections
is a point of primary importance.

Due to the interaction with environment, each ``doorway'' resonance
state excited in the structure via external channels gets fragmented
onto a large number $\sim\Gamma_s/d$ (the
spreading width $\Gamma_s$ characterizes the strength of the
coupling to the environment when $d$ is the single-quasi-particle
mean level spacing) of very narrow resonances \cite{Sokol10}.
Only the cross sections averaged over the fine $(\thicksim d)$ structure scale are
observable. Due to such an averaging the doorway resonance states
are damped not only because of escape through such channels but
also due to the ulterior population of the relatively long-lived environmental
states. As a result, transmission of an electron with a given incoming
energy $E_{in}$ through the structure turns out to be an incoherent
sum of the flow formed by the interfering damped doorway
resonances and the retarded flow of the particles re-emitted into the
structure by the environment. Being delayed, the returning electrons
do not interfere with those that escape directly through the external
leads.

The temporal characteristics of the transport are described in detail by the $M\times M$ Smith time delay matrix $Q=-iS^{\dag}dS/dE=Q^{(s)}+Q^{(e)}$
that consists of two contributions $Q^{(s,e)}$. They correspond respectively to the
(modified due to the interaction with the background) time delay within the dot and
additional delay because of the temporary transitions into the background.

We suppose the temperature of the environment to be zero whereas
the energy of the incoming particle $E_{in}$ can be close to or
somewhat above the Fermi surface of electrons in the environment.
Therefore, though the number of the particles is definitely conserved
in each individual event of transmission, there exists a probability
that some part of the electron's energy can be absorbed due to
environmental many-body effects. The both decoherence and
absorption phenomena can be naturally treated within the framework of a
unit microscopic model based on the general theory of the resonance
scattering. Both these effects are controlled by the only parameter:
the spreading width of the doorway resonances.

If the energy $E_{in}$ noticeably exceeds the environment's Fermi
surface and the doorway resonances overlap, the random matrix
approach becomes relevant and ensemble averaging in the doorway
sector is appropriate. Such an averaging, being equivalent to the
energy averaging over the doorway scale $D$, suppresses all
interference effects  save the elastic enhancement phenomenon. The latter
is a direct consequence of the time reversal symmetry and manifests itself
in the so called {\it weak localization effect}. However the energy
absorption in the environment violates this symmetry and suppresses the weak
localization.

\section{Particulars}
\label{sec:Particulars}

\subsection{An example of chaotic classical dynamics}
\label{sec:exmplClassChaoDyn}
    As an instructive and typical example of chaotic classical system
we consider below a periodically kicked quartic nonlinear oscillator
\begin{equation}\label{H_c}
 H(\alpha^{*},\alpha;t)=H^{(0)}+H^{(k)},
\end{equation}
where the unperturbed Hamiltonian function reads
\begin{equation}\label{H^0_c}
  H^{(0)}(\alpha^{*},\alpha)=
    \frac{p^2}{2m}+\frac{\omega^2_0}{4}\tan^2(\sqrt{2m}\, x)=
    \omega_0\,|\alpha|^2+|\alpha|^4 ,
\end{equation}
and the time-dependent perturbation
\begin{equation}
  H^{(k)}(\alpha^{*},\alpha;t)=g(t)=
    g_0\sum_{\tau}\delta(t-\tau)\,(\alpha^*+\alpha) .
\end{equation}
constitutes a sequence of periodic kicks with strength $g_0$ that
are acting at the instances $\tau=\pm 1, \pm 2, ...\,$.

There are two convenient choices of the canonical variables in this
case: $I, \theta$ or $\alpha, i\alpha^*$  related by the canonical
transformation
$\alpha=\sqrt{I}\,e^{-i\theta}$, $\alpha^*=\sqrt{I}\,e^{i\theta}$.
The action-angle variables $I, \theta$ are ordinary used in the
classical considerations. On the other hand, the advantage of
$\alpha$-variables is that they are directly related to the quantum
creation-annihilation operators ${\hat a}^{\dagger}, {\hat a}$.
The action-angle variables satisfy along a given phase trajectory two
coupled nonlinear equations \cite{Sokol84}:
\begin{eqnarray}
  I(t)&=&\left|{\sqrt{{\overset{\circ}I}}
       +i\int_0^t d\tau
           g(\tau)e^{i[\theta(\tau)-\overset{\circ}\theta]}}\right|^2\\
  \theta(t)&=&\int_0^t d\tau[\omega_0+2I(\tau)]
\end{eqnarray}
(here and in what follow we mark the initial values by the overset
circle).    A great majority of trajectories becomes exponentially
unstable and,  the corresponding motion is globally chaotic when the
strength of the kicks exceeds 1: $|g_0|>1$. Under this condition the
phase correlations decay with time exponentially fast \cite{Sokol07},
\begin{equation}\label{ClasCorrDecay}
  \Big|\int d \overset{\circ}I\,d\overset{\circ}\theta\,
     W^{(c)}(\overset{\circ}I\,,\overset{\circ}\theta;t=0)\,
     e^{i\left(\theta(t)-\overset{\circ}\theta\right)}\Big|^2=
  \exp(-t/\tau_c),
\end{equation}
where the characteristic time $\tau_c\sim 1/\Lambda$ is directly
connected to the Lyapunov exponent $\Lambda$. Here
$W^{(c)}(I\,,\theta;t=0)$ is the initial probability distribution in the
phase space. This exponential decay of phase correlations is an
universal fingerprint of the classical dynamical chaos.

In fact, almost all trajectories are alike when the motion is globally
chaotic and, actually, only behavior of manifolds of them is of the real
interest. Therefore the phase space methods appear to be most
relevant in the chaotic regime. The classical phase-space distribution
function $W^{(c)}(\alpha^{*},\alpha;t)$ satisfies the linear Liouville
equation
\begin{equation}\label{L_c_eqn}
  i\frac{\partial}{\partial t}\,W^{(c)}(\alpha^*,\alpha;t)=
  {\cal\hat L}_c(t)\,W^{(c)}(\alpha^*,\alpha;t)
\end{equation}
with the unitary Liouville operator ${\cal\hat L}_c(t)$ that is a sum
of two, stationary and time-dependent, parts:
${\cal\hat L}_c(t)={\cal\hat L}_c^{(0)}+{\cal\hat L}^{(k)}(t)$.
The first, unperturbed part reads \cite{Sokol08a}
\begin{equation}\label{L_0_c}
  {\cal\hat L}_c^{(0)}=
    \left(\omega_0+2|\alpha|^2\right)
    \left(\alpha^*\frac{\partial}{\partial\alpha^*}-
      \alpha\frac{\partial}{\partial\alpha}\right),
\end{equation}
where the operator
$ \left(\alpha^*\frac{\partial}{\partial\alpha^*}-
   \alpha\frac{\partial}{\partial\alpha}\right)
  =-i\frac{\partial}{\partial\theta}$
formally coincides with the quantum-mechanical angular momentum
operator $\frac{1}{\hbar}{\hat L}_z$. In fact, this operator describes
rotation in the phase-space around the origin with a local angular
velocity $\left(\omega_0+2|\alpha|^2\right)$.

The time-dependent perturbation (kick) operator
\begin{equation}\label{L_k}
  {\cal\hat L}^{(k)}(t)=
  g_0\sum_{\tau}\delta(t-\tau)
    \left(\frac{\partial}{\partial\alpha^*}-
          \frac{\partial}{\partial\alpha}\right)
\end{equation}
describes sequence of instant shifts by the distance $g_0$ in the
$\alpha$-plane. The alternating twists and shifts develop
unpredictably complicated pattern of the density distribution
$W^{(c)}(\alpha^*,\alpha;t)$ when the perturbation strength
constant $|g_0|>1$ \cite{Chiri79}. It is of primary importance here
that the unperturbed part ${\cal\hat L}_c^{(0)}$ of the Liouville
operator has \textit{continuous} spectrum of eigenvalues. As a
consequence, the classical phase distribution is structuring
exponentially fast on finer and finer scale during chaotic evolution.
Such a behavior is the paramount property of the classical dynamical
chaos.

Fourier analysis in the phase plane provides a natural tool for
elucidating the process of this structuring. Taking into account
periodicity of the distribution function $W^{(c)}(\alpha^*,\alpha;t)$
with respect to the angle $\theta$,
\begin{equation}\label{Four_repr}
W^{(c)}(I,\theta;t)=
  \frac{1}{2\pi}\sum_{m=-\infty}^{\infty}
                   W_m^{(c)}(I;t)\,e^{i m\theta}\,,
\end{equation}
the most simple and efficient idea is just to follow the upgrowth of the
number of its $\theta$-harmonics. It can be easily quantified with the
help of notion of the Peres fidelity. We suppose for simplicity that the
initial distribution has been isotropic,
$W_m^{(c)}(I;0)=0$ if $m\neq 0$
and let the system to evolve autonomously during some time $t_r$.
At this moment, we probe the system with the help of an instant weak
perturbation $\xi I\delta(t-t_r)$ that produces rotation in the phase
plane by some angle $\xi$. To simplify slightly  the subsequent
formulae we suggest the parameter $\xi$ to be a Gaussian random
variable. The probe results then in the instant change of the
distribution
$W^{(c)}(I,\theta;t_r-0)\rightarrow W^{(c)}(I,\theta+\sigma;t_r+0)$
where the parameter $\sigma$ sets the mean squared strength
of the perturbation. The Peres fidelity defined in this case as
\begin{equation}\label{F_c_sen}
F_{(sen)}(\sigma;t_r)=\frac{\int
d^2\alpha\,W^{(c)}\left(\alpha^*,\alpha;t_r\right)
W^{(c)}\left(\sigma\big|\alpha^*,\alpha;t_r+0\right)}
{\int d^2\alpha\,[W^{(c)}\left(\alpha^*,\alpha;t_r\right)]^2}
=\sum_{m=0}^{\infty} e^{-\frac{1}{2}\sigma^2 m^2}
     \mathcal{W}^{(c)}_m(t_r)
\end{equation}
characterizes sensitivity of the motion to an external influence.

The quantities $\mathcal{W}^{(c)}_m(t)$
\begin{equation}\label{m_weights_c}
\mathcal{W}^{(c)}_m(t)=(2-\delta_{m0})\,
\frac{\int_0^{\infty}d I\,\Big|W^{(c)}_m(I;t)\Big|^2}
{\sum_{m=-\infty}^{+\infty}\,\int_0^\infty dI |W^{(c)}_m(I;t)|^2}
\end{equation}
that are expressed in the terms of the Fourier harmonics and satisfy
the normalization condition
\begin{equation}\label{W_norm_c}
\sum_{m=0}^{\infty}\mathcal{W}^{(c)}_m(t)=1,
\end{equation}
can naturally be interpreted in the probabilistic manner as the
weights of the corresponding $\theta$-harmonics. Therefore we can
define in the spirit of the linear response approach the mean number
of them $\langle |m|\rangle_t=\sqrt{\langle m^2\rangle_t}$ at a
moment of time $t$ with the aid of the relation
\begin{equation}\label{Mean_m_c}
\langle m^2\rangle_t=
   \sum_{m=0}^{\infty} m^2\, \mathcal{W}^{(c)}_m(t)=
   -\frac{d^2 F_{(sen)}(\sigma;t)}{d\sigma^2}\big|_{\sigma=0}\,.
\end{equation}
The number $\langle |m|\rangle_t$ can serve as a convenient
quantitative measure of complexity of a phase space distribution at a
given moment of time $t$. Due to exponential instability of classical
dynamics the number of harmonics proliferates exponentially
$\langle |m|\rangle_t\propto e^{t/\tau_c}$ \cite{Sokol08a}.

Let us suppose now that the motion has been reversed at the
moment $t_r$, immediately following after the
time of probing. In view of the fact that the
Liouville evolution operator is unitary one we can transform the
expression ({\ref{F_c_sen}}) into the form
\begin{equation}\label{F_c_rev}
F_{(sen)}(\sigma;t_r)=
  \frac{\int d^2\alpha\,W^{(c)}\left(\alpha^*,\alpha;0\right)
  W^{(c)}\left(\sigma\big|\alpha^*,\alpha;t_r+0,0\right)}
  {\int d^2\alpha\,[W^{(c)}\left(\alpha^*,\alpha;0\right)]^2}
=F_{(rev)}(\sigma;t_r)\equiv F(\sigma;t_r)\,.
\end{equation}
Transformed in this form, fidelity measures the overlap of the
isotropic initial distribution function and the distribution
$W^{(c)}\left(\sigma\big|\alpha^*,\alpha;t_r+0,0\right)$
that, after being changed by the probing perturbation at the moment
$t_r$, has reverted to the initial moment $t=0$. This formula
connects the reversibility of the motion with sensitivity to external
perturbations or, by other words, with complexity of the distribution
function at the reversal moment $t_r$.

\subsubsection{Deterministic diffusion and onset of irreversibility}
\label{sec:Diffusion&classIrrevers}
    The phenomenon of the so called deterministic diffusion is one of the
simplest manifestation of the classical dynamical chaos. The mean
value of the action $I(t)$ at any given time $t$ is calculated as follows:
\begin{equation}\label{I_c_mean}
\langle I\rangle_t=\int_0^{\infty}dI\,I\,W^{(c)}_0(I;t)=
  \int_0^{\infty}dI\,I\,\overline{W^{(c)}_0(I;t)}=
  \langle I\rangle_0+g_0^2\,t
\end{equation}
see, e.g. \cite{Chiri79,Sokol84a,Abarb82}. The averaging is performed in two steps
here. We have smoothed first
the very irregular function $W^{(c)}_0(I;t)$ over a small interval
surrounding a fixed value $I$ within which the factor $I$ does not
appreciably change. Successive $I$-integration with the ``coarse
grained'' distribution $\overline{W^{(c)}_0(I;t)}$ results in the
diffusive increase of the mean action. Similarly, we can also calculate
the moments $\langle I^k\rangle_t,\, k=2, 3,...,k_t$. The longer is the
duration $t$ of the evolution the larger the power $k_t$ at which the
two-step integration procedure is still valid.

It is obvious that, formally, the deterministic diffusion
(\ref{I_c_mean}) is perfectly time-reversible.
However the exponential instability makes this statement impractical.
Even a very weak external noise destroys reversibility and turns the
motion into irreversible process.

To illustrate this statement, we add to our Hamiltonian function a new
term
\begin{equation}\label{noise_c}
H^{(noise)}(\alpha^*,\alpha;t)=
     |\alpha|^2\sum_{\tau}\xi_{\tau}\,\delta(t-\tau),
     \quad \langle\xi_{\tau}\rangle=0,\,
     \langle\xi_{\tau}\xi_{\tau'}\rangle=\sigma^2\delta_{\tau\tau'}\,.
\end{equation}
that describes a stationary Gaussian noise with the strength
parameter $\sigma$. Each period of unperturbed evolution is
immediately followed by a phase plane rotation by a random angle
$\xi$. Averaging over the noise realizations sets up the coarse grained
distribution function
\begin{equation}\label{c_g_W}
  W^{(c)}(\sigma|I,\theta;t)=
                \overline{W^{(c)}(\{\xi\}|I,\theta;t)}^{(noise)}.
\end{equation}
The averaging suppresses Fourier harmonics with respect to the both
canonical variables thus extenuating irregular oscillations of the
distribution. They are perfectly smoothed away in the limit of very
strong noise
$\sigma\rightarrow\infty$ :
$W^{(c)}_{|m|\geqslant 1}(\infty|I,;t)=0$
and
\begin{equation}\label{Inf_noise}
  W^{(c)}_{0}(\infty|I;t)=\frac{1}{\langle I\rangle_0+g_0^2\,t}
  \exp\left(\frac{I}{\langle I\rangle_0+g_0^2\,t}\right).
\end{equation}
In this case the fidelity (\ref{F_c_rev}) equals
$F_{(rev)}(\infty|t_r)=e^{-\frac{t_r}{\tau_c}}\,.$
Indeed, the backward evolution starts with the isotropic distribution
(\ref{Inf_noise}) at the moment $t_r$ and develops
$m(t_r)\propto e^{t_r/\tau_c}$
$\theta$-harmonics by the time $t=0$.
Thus the overlap with the initial isotropic distribution
$W^{(c)}\left(\alpha^*,\alpha;0\right)$ is exponentially small. The
standard diffusion takes place in the backward evolution as well.
Under influence of the noise of a moderate level $\sigma$ the
backward diffusion is delayed for some time $\tau_d\propto\ln\sigma$
\cite{Iked95}
during which the system partly recovers its preceding states. But
after that the diffusion recommences.

\subsection{Quantum evolutions of classically chaotic system}
\label{sec:QuantumEvolutionOfChaos}
    The Hamiltonian of the quantum counterpart of the classical oscillator
considered above is immediately obtained by the substitutions:
 $\alpha\Rightarrow\sqrt{\hbar}\,{\hat a}$,
 $\alpha^*\Rightarrow\sqrt{\hbar}\,{\hat a}^{\dag}.$
\begin{equation}
 {\hat H}({\hat a}^{\dag},{\hat a};t)=\hbar\omega_0 {\hat n}+
  \hbar^2{\hat n}^2+g_0\sum_{\tau}\delta(t-\tau)\,
  \sqrt{\hbar}\,({\hat a}^{\dag}+{\hat a}).
\end{equation}
(In the chosen units all parameters: $\hbar$, $\omega_0$, $g_0$
are dimensionless.)

The quantum evolution
$\hat\rho(t)=\hat{\cal U}(t)\hat\rho(0)\hat{\cal U}^{\dag}(t)$
of the density matrix is described by the unitary operator
$\hat{\cal U}(t)$ that is $t$ successive repetitions of the one period
Floquet operator $\hat U$: $\hat{\cal U}(t)={\hat U}^t$,
the latter being a sequence of instant change of the excitation
number (coherent state displacement operator) and subsequent free
rotation in the phase plane induced by the time-independent
Hamiltonian
${\hat H}^{(0)} ({\hat a}^{\dag},{\hat a})=
    \hbar\omega_0 {\hat n}+\hbar^2{\hat n}^2$:
\begin{equation}
  \hat U=e^{-\frac{i}{\hbar}
         \hat{H}^{(0)}}\hat D\left(i\frac{g_0}{\sqrt\hbar}\right)
\label{eq:U}
\end{equation}
The Wigner function (\ref{Wfunc}) satisfies the quantum Liouville
equation:
\begin{equation}\label{L_q_eqn}
i\frac{\partial}{\partial t}\,W(\alpha^*,\alpha;t)=
     {\cal\hat L}_q(t)\,W(\alpha^*,\alpha;t)\\
\end{equation}
where the quantum Liouville operator is, similar to the classical one,
the sum
${\cal\hat L}_q(t)={\cal\hat L}_q^{(0)}+{\cal\hat L}^{(k)}(t)$.
The only, but of primary importance, distinction from the classical
case is appearance of a new second-derivative term in the rotation
operator
\begin{equation}\label{L_0_q}
 {\cal\hat L}_q^{(0)}=\left(\omega_0-\hbar-\frac{1}{2}\hbar^2
 \frac{\partial^2}{\partial\alpha^*\partial\alpha}+2|\alpha|^2\right)
 \left(\alpha^*\frac{\partial}{\partial\alpha^*}-
 \alpha\frac{\partial}{\partial\alpha}\right).
\end{equation}
The combination
$-\frac{1}{2}\hbar^2\frac{\partial^2}{\partial\alpha^*\partial\alpha}
+2|\alpha|^2$
formally coincides with the quantum Hamiltonian of a 2D isotropic
linear oscillator. As a result, the spectrum of the operator
(\ref{L_0_q}) becomes discrete, in contrast to his classical
counterpart. This important fact is a manifestation of {\it the
 quantization of the phase space}.

Below we will consider evolution of an initially isotropic and,
generally, mixed state. If we choose the density matrix in the form
\begin{equation}\label{In_rho}
  {\hat\rho}(0)=
    \frac{\hbar}{\Delta+\hbar}
    \sum_{n=0}^{\infty}\left(\frac{\Delta}{\Delta+\hbar}\right)^n
    |n\rangle\langle n|
\end{equation}
the corresponding Wigner function
\begin{equation}\label{W_q_in_Delta}
  W(\alpha^*,\alpha;0)=
    \frac{1}{\Delta+\hbar/2}\,e^{-\frac{|\alpha|^2}{\Delta+\hbar/2}}.
\end{equation}
is a Poissonian distribution with respect to the action variable
$I=|\alpha|^2$.
In particular, in the case $\Delta=0$ this state turns into the pure
ground state ${\hat\rho}(0)=|0\rangle\langle 0|$ whose Wigner
function
$W(\alpha^*,\alpha;0)=
\frac{2}{\hbar}\, e^{-\frac{2|\alpha|^2}{\hbar}}$
occupies the minimal quantum cell $\hbar/2$.

As compared with the classical Liouville equation the quantum one
\begin{equation}
  i\frac{\partial}{\partial t}\,W(\alpha^*,\alpha;t)=
    {\cal\hat L}_c\,W(\alpha^*,\alpha;t)
-\frac{1}{2}\hbar^2\frac{\partial^2}{\partial\alpha^*\partial\alpha}
    W(\alpha^*,\alpha;t);
\end{equation}
contains an additional proportional to $\hbar^2$ term with higher
(second) order derivative. Even being initially very small, this term
can be neglected only, because of the classical exponential instability,
during a short time interval which is called the Ehrenfest time
$t_E=\tau_c\,\ln\frac{2\langle I\rangle_{t_E}}{\hbar}$. After this
interval the quantum effects dominate and difference between the
two dynamics becomes crucial. Indeed, meshing of the Wigner
function stops quite soon on account of the quantization of the phase
space.

\subsection{Stability and reversibility vs complexity of quantum states.}
\label{sec:quStabRev&Complexity}
    Relying upon the analogy mentioned above between the classical
phase space distribution function on the one hand and the quantum
Wigner function on the another we will confront below the typical
features of the two corresponding "chaotic" dynamics. Quantization of
the phase space implies a much simpler structure of the Wigner
function than that of the corresponding classical distribution. The
Liouvillian phase space approach that we use allows us to measure
complexity of this function using the same tools that have been
utilized in the classical case considered in subsection \ref{sec:exmplClassChaoDyn}.

\subsubsection{Response to a single probe}
\label{sec:SingleProbe}
\begin{figure}[!h]
\centering
\includegraphics[width=80mm,
  keepaspectratio=true]{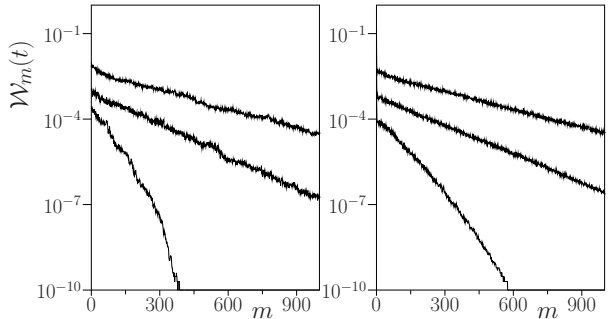}
\caption{Distribution of harmonics ${\cal W}_m(t)$ as a function
of $m$, at different times $t=10, 30$ and $50$ from bottom to top
(these curves are scaled by a factor 0.01, 0.1 and 1, respectively).
Left and right panels correspond to pure ($\Delta=0$) and mixed
($\Delta=25$) initial states, see eq.(\ref{In_rho}). Other parameters
of simulations are $g_0=2$, $\hbar=1$.}
\label{fig:quWm}
\end{figure}

    In much the same fashion as in eqs. (\ref{F_c_sen}-\ref{F_c_rev}),
the response of a quantum system to an instant probe at a moment of
time $t$ is measured with the help of the quantum Peres fidelity
(\ref{Fidelity}) defined by the relation
\begin{equation}\label{F_q}
F(\sigma|t)=\frac{\mathrm{Tr}\left[{\hat\rho(t)}\,
     {\hat\rho(\sigma|t)}\right]}{\mathbb{P}(t)}=
      \sum_{m=0}^{\infty}
                           e^{-\frac{1}{2}\sigma^2 m^2}\mathcal{W}_m(t)
\end{equation}
where the quantities
\begin{equation}\label{calW_m_q}
 \mathcal{W}_m(t)=
     \frac{(2-\delta_{m0})}{\mathbb{P}(t)}\sum_{n=0}^{\infty}
     \Big|\langle n+m\big|{\hat\rho}(t)\big|n\rangle\Big|^2,\quad\,
     \sum_{m=0}^{\infty}\mathcal{W}_m(t)=1
\end{equation}
are direct quantum analogs of the classical weights
(\ref{m_weights_c}) of the $\theta$-harmonics. Indeed, they can be
expressed \cite{Sokol08a,Sokol08} in the terms of Fourier amplitudes
of the Wigner function by the formulae:
\begin{equation}\label{m_weights_q}
  \mathcal{W}_m(t)=
  \frac{(2-\delta_{m0})}{\mathbb{P}(t)}\,
  \hbar\int_0^{\infty}d I\,\Big|W_m(I;t)\Big|^2
\end{equation}
(see Fig.\ref{fig:quWm}).

\begin{equation}\label{Purity_q}
  \mathbb{P}(t)=
    \mathrm{Tr}\left[{\hat\rho}^2(t)\right]=
    \sum_{m=-\infty}^{+\infty}\,
          \int_0^\infty dI |W_m(I;t)|^2=\mathbb{P}(0).
\end{equation}
These relations differ from eqs. (\ref{m_weights_c}, \ref{W_norm_c})
only by replacement $W_m^{(c)}(I;t)\Rightarrow W_m(I;t)$ of the
Fourier amplitudes of the classical distribution function by those of
the quantum Wigner function. It is worth noting that the phases of
the off-diagonal matrix elements of the density matrix at the moment
$t$ fall out of the weights (\ref{calW_m_q}). Nevertheless such
phases do play a certain, though not significant (see below), role
during the preceding evolution.

\subsubsection{Complexity of a quantum state.}
\label{sec:ComplexityQuState}
\begin{figure}
\centering
\includegraphics[width=8.cm,angle=0
]{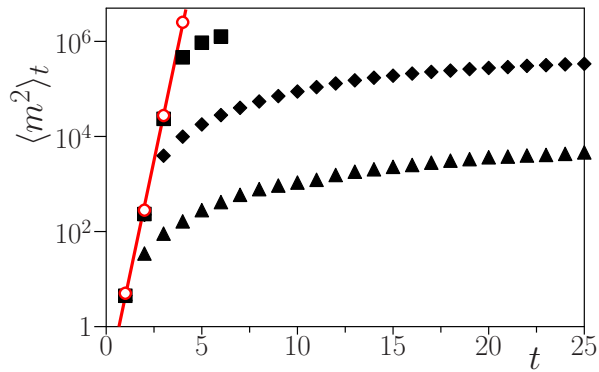} 
\caption{$\langle m^2\rangle_t$ before and after the Ehrenfest time.
Triangles, diamonds and squares: $\hbar = 1,\, 0.1$ and $0.01$
respectively. Classical dynamics is shown by empty red circles,
the gray line presents an exponential fit.
The initial phase area {\it holds constant} $1/2$ in all 3 cases.
The kick strength used here is $g_0=1.5$.}
\label{fig:Wquantclass}
\end{figure}

    Just as it has been in the case of a classical phase space distribution,
complexity of a quantum state can be characterized by the number
$\langle |m|\rangle_t$ of $\theta$-harmonics of the Wigner function.
The corresponding weights are given now by eqs. (\ref{m_weights_q},
\ref{Purity_q}) and again
\begin{equation}\label{mean_m_q}
  \langle m^2\rangle_t=
    \sum_{m=0}^{\infty} m^2\,\mathcal{W}_m(t)=
    -\frac{\rm d^2 F(\sigma;t)}{\rm d\sigma^2}\Big|_{\sigma=0}.
\end{equation}
It should be emphasized that the chosen measure is {\it equally valid}
in the quantum as well as classical cases. This allows direct
comparison of the main features of the both dynamics. In particular,
whereas the number of harmonics of the classical distribution
function increases, due to the exponential instability, exponentially
during the {\it whole}  time of evolution, in the quantum case the
exponential regime is, generally speaking, restricted to the Ehrenfest
time interval. These statements are illustrated in the
Fig.\ref{fig:Wquantclass} where the dependence of the quantity
$\langle m^2\rangle_t$ on the time is shown for a set of different
values of the effective Plank's constant.

The Plank's constant plays here a twofold role: on the one part
it fixes the phase volume of the elementary quantum cell that is
occupied in the phase space by the pure ground state
${\hat\rho}(0)=|0\rangle\langle 0|$, and, on the other part, it governs
the dynamics via the evolution equation. In the
Fig.\ref{fig:Wquantclass} we keep constant the size 1/2 of the initial
Wigner distribution (\ref{W_q_in_Delta}) by choosing
$\Delta=1/2-\hbar/2$. The initial state is pure ($\Delta=0$) when
$\hbar=1$ (triangles) but becomes more and more mixed
($\Delta\approx 1/2\gg\hbar$) in the cases $\hbar=0.1$ (diamonds)
and $\hbar=0.01$ (squares) correspondingly. It is clearly seen that
the smaller is the value of {\it dynamical} Plank's constant the longer
the classical exponential regime lasts. By other words, the mixing of
the initial state suppresses the quantum interference effect and
restores the classical behavior.

\subsubsection{Information entropy vs von Neumann entropy}
\label{sec:Entropies}
    The probabilistic meaning of the quantities $\mathcal{W}_m(t)$
allows us to introduce the Shannon or {\it information entropy}:
\begin{eqnarray}\label{Inf_entropy}
{\cal I}(t)&=&
  -\sum_{m=0}^{\infty}\mathcal{W}_m(t)\,\ln\mathcal{W}_m(t)\\
  &=&
    \left\{
      \begin{array}{ll}
        0, & t=0,\\
        \ln\langle |m|\rangle_t+1-\frac{\ln 2}{2}+O(1/\langle |m|\rangle_t), \qquad &\langle |m|\rangle_t\gg 1.\\
      \end{array}
    \right.
\end{eqnarray}
This is another, though equivalent, possible way to characterize the
complexity of a quantum state. Such a choice turns out to be even
more convenient for our further purposes. This entropy starts from
zero (the initial state has no harmonics but zero), increases linearly
(with a slope defined by the classical characteristic time $\tau_c$)
during the Ehrefest time and then slows down to the quantum
logarithmic regime. Since the number of harmonics at sufficiently
large time weakly depends on the peculiar properties of the initial
state, being practically the same for pure and mixed ones, this
entropy is practically insensitive to quantum correlations.

On the contrary, the invariant von Neumann entropy
\begin{equation}
  {\cal S}(t)=
  -\mathrm{Tr}  \left[{\hat\rho}(t)\,\ln{\hat\rho}(t)\right]={\cal S}(0)
\end{equation}
is {\it perfectly} sensitive to quantum correlations (hence the name
``correlational'' \cite{Sokol98}). This entropy does not depend on
time as long as the evolution remains unitary and equals zero
${\cal S}(t)=0$ when the state is pure. The coherence and quantum
correlations can be destroyed only in the presence of an persistent
external noise or during the process of preparation of the initial state
(see below).

\subsubsection{Persistent noise.}
\label{sec:Noise}
\begin{figure}[!h]
\centering
\includegraphics[width=80mm, keepaspectratio=true]
{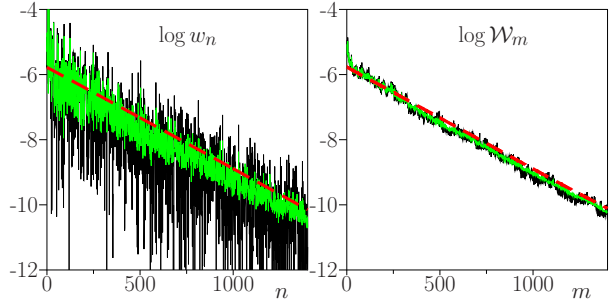}
\caption{Probability distributions
(\ref{eq:rho_xi}), (\ref{calW_m_q_xi})  at the moment $t=80$
with no noise ($\sigma=0$, thin black lines), weak noise
($\sigma=0.001$, thin green lines) and strong noise ($\sigma =1$,
thick dashed red line). In the two latter cases the distributions are
averaged over $10^3$ realizations.  The initial state is pure ($\Delta=0$, see eq.(\ref{In_rho})). Other parameters
of simulations are $g_0=2$, $\hbar=1$.}
\label{fig:ns_wn_Wm}
\end{figure}

    The stationary noise is described in our model by the Hamiltonian
operator
\begin{equation}\label{noise_q}
  {\hat H}^{(noise)}({\hat a}^{\dag},{\hat a};t)=
          \hbar\,{\hat n}\sum_{\tau}\xi_{\tau}\,\delta(t-\tau),
          \quad \langle\xi_{\tau}\rangle=0,\,
          \langle\xi_{\tau}\xi_{\tau'}\rangle=
          \sigma^2\delta_{\tau\tau'}\,.
\end{equation}
As a result, the evolution operator takes the form
\begin{equation}\label{U(ksi,t)}
\hat{\cal U}(\{\xi\};t)=
\prod_{\tau=1}^{\tau=t}\left[e^{-i\xi_{\tau}{\hat n}}\,\hat U\right].
\end{equation}
This operator remains unitary for any fixed noise realization $\{\xi\}$
(history). Accordingly, a pure initial state remains pure during
the whole time of evolution. At a running moment $t$, the
excitation of the oscillator and the degree of
anisotropy of the Wigner function are characterized by the
probability distributions \cite{Sokol09}
\begin{equation}
 w_n({\xi}; t)=\langle n | \hat{\rho}({\xi}; t) |n\rangle,
 \label{eq:rho_xi}
\end{equation}
and
\begin{equation}\label{calW_m_q_xi}
 \mathcal{W}_m(t)=
     \frac{(2-\delta_{m0})}{\mathbb{P}(t)}\sum_{n=0}^{\infty}
     \Big|\langle n+m\big|{\hat\rho}({\xi}t)\big|n\rangle\Big|^2,
     \quad\,
     \sum_{m=0}^{\infty}\mathcal{W}_m(t)=1
\end{equation}
where the density matrix
\begin{equation}
 \hat{\rho}({\xi};t)=
   \hat{\cal U}(\{\xi\};t)\hat\rho(t=0)\hat{\cal U}^{\dagger}(\{\xi\};t)
\end{equation}
is defined for some fixed noise realization $\{\xi\}$. The initial
state can be chosen to be the ground one
$ \hat\rho(t=0)=\big|0\rangle\langle 0\big|$, because after a few first
kicks the distributions (\ref{eq:rho_xi},\ref{calW_m_q_xi}) aquire
a practically general exponential form (see Fig.\ref{fig:ns_wn_Wm}).
Averaging over noise realizations keeps the
slopes of these distributions
unchanged but kills the wild fluctuations around their regular
exponential decay. In the limit of strong noise $\sigma\gg1 $
(see thick dashed lines in Fig.\ref{fig:ns_wn_Wm}) the
noise acts as a ``coarse graining'' that reproduces the behavior of
the corresponding classical distributions \cite{Sokol09}.
There exist "self-averaging" quantities
like mean excitation number $\langle n\rangle_t$ or the mean
number $\langle |m|\rangle_t$ of $\theta$-harmonics that do not
depend, in fact, on the noise realization \cite{Sokol09}
\begin{equation}
  \langle n(\{\xi\};t)\rangle=n(\sigma;t),\,\,\,
  \langle m^2(\{\xi\};t)\rangle=
    m^2(\sigma;t)=n(\sigma;t)\left(n(\sigma;t)+1\right).
\end{equation}

\begin{figure}[!h]
\centering
\includegraphics[width=80mm,
  keepaspectratio=true]{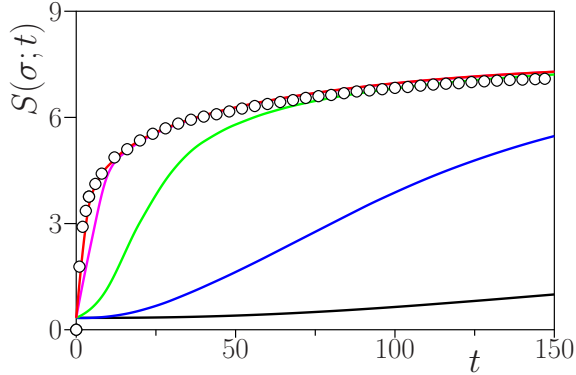}
\caption{Von Neumann entropy:
 $\sigma= (0.125,1, 8, 64, 512)\cdot 10^{-3}$,
solid (black, blue, green, magenta, red) lines from bottom to top.
 Circles: information entropy
 ${\cal I}(t);\,\,\,{\cal S}(\sigma;t>t_{(dec)}) \to {\cal I}(t)\,$.
 In these simulations we use $g_0=2$, $\hbar=1$.}
\label{fig:two_entropies}
\end{figure}

On the contrary, the quantum Peres fidelity defined as
\begin{equation}\label{Fid_q}
F(\{\xi\}; t)=\frac{\mathrm{Tr}
      \left[{\hat\rho(t)}\, {\hat\rho(\{\xi\};t)}\right]}{\mathbb{P}(t)}.
\end{equation}
is not a self-averaging quantity and wildly fluctuates from one noise
history to
another. Therefore averaging over all possible noise realizations is
necessary to obtain a reasonably simple and adequate measure:
\begin{equation}\label{Fid_q_av}
 F(\sigma;t)=\overline{F(\{\xi\};t)}^{\{\xi\}}
    =\mathrm{Tr}
       \left[{\hat\rho(t)}\,{\hat\rho^{(av)}(\sigma;t)}\right] .
\end{equation}
This procedure brings into consideration the average density matrix
$\rho^{(av)}(t)$ whose one step evolution is described by the
transformation
\begin{equation}\label{1step}
  \langle n'|{\hat\rho^{(av)}(\sigma;\tau)}|n\rangle=
      e^{-\frac{1}{2}\sigma^2(n'-n)^2}\,
      \langle n'|{\hat U}{\hat\rho^{(av)}(\sigma;\tau-1)
      {\hat U}^{\dag}}|n\rangle.
\end{equation}
The noise suppresses off-diagonal matrix elements of the density
matrix thus gradually cutting down the number of harmonics of the
corresponding Wigner function. The evolution is not unitary
anymore. The latter entails state mixing, loss of memory on the
initial state and suppression of the quantum interference. These
effects show up in the behavior of the von Neumann entropy defined
in the terms of the averaged density matrix  as
\begin{equation}\label{N_entropy_ev}
 {\cal S}(\sigma;t)= -\mathrm{Tr}
      \left[{\hat\rho}^{(av)}(\sigma;t)\,\ln{\hat\rho}^{(av)}(\sigma;t)
  \right].
\end{equation}
The Fig.\ref{fig:two_entropies} illustrates this behavior in
comparison with that of the information Shannon entropy
${\cal I}(t)$ (see (\ref{Inf_entropy})). The entropy
${\cal S}(\sigma;t)$ increases the faster the larger is the level of
noise (the full lines from bottom to top) and approaches the
information Shannon ${\cal I}(\sigma;t)$ entropy (circles) from
below. At some time $t_{(dec)}$, both the entropies coincide and go
after that together. All coherent effects are washed away by this
time. Therefore the entropy (\ref{N_entropy_ev}) is suitable
{\it for tracing the gradual loss of the quantum coherence.}

The decoherence time $t_{(dec)}$ can be estimated as
\cite{Sokol09}
\begin{equation}
  t_{(dec)}(\sigma)\thicksim\sqrt{\frac{\hbar}{\sigma^2 D}}
\end{equation}
where $D$ is the classical diffusion coefficient. Henceforth the
system occupies the maximal phase area accessible at the running
value of excitation thus reaching a sort of equilibrium. Finally, the
system expands "adiabatically", the both entropies being almost
constant. The further evolution turns out to be \textit{Markovian}
\cite{Sokol09}.

\subsubsection{Time independent perturbation. Mixed initial state.}
\label{sec:TimeIndependPertirb}
    As it has already been noted above, the quantum interference turns
out to be somewhat suppressed if the evolution started from a mixed
initial state. We consider below  another kind of decoherence that
takes place in the case of a time-independent perturbation.
\begin{equation}
  \hat{H}_V=\hat{H}^{(0)}+\varepsilon {\hat V},
\end{equation}
where the unperturbed Hamiltonian describes, as before, dynamics
of the classically chaotic nonlinear oscillator. The perturbed and
unperturbed motions are juxtaposed by means of the unitary
Loschmidt echo operator ${\hat f}(t)$ \cite{Sokol07}
\begin{eqnarray}
  {\hat f}(t)&=&{\hat U}^{\dag}(t){\hat U}_V(t)\\
  {\hat f}(t)&=&T\exp\left[-i\frac{\varepsilon}{\hbar}\int_0^t d\tau
       {\hat{\cal V}}(\tau)\right]\\
  {\hat{\cal V}}(\tau)&=&{\hat U}^{\dag}(t)\,{\hat V}{\hat U}(t).
\end{eqnarray}
In spite of the fact that the perturbation ${\hat{\cal V}}(\tau)$
evolves chaotically, the quantum coherence {\it is in no way
spoiled} as long as the initial state is pure. More than that, one
might think that even if the initial state is an incoherent mixture
quantum coherence can be rapidly generated by producing complex
off diagonal matrix elements during dynamical evolution.
Nevertheless, we will show below that if the system is classically
chaotic and the evolution \emph{starts from a wide incoherent
mixed state}, then the initial incoherence persists due to the
intrinsic classical chaos so that the quantum phases remain
irrelevant \cite{Sokol07}.

In the case of a pure initial state
${\hat\rho}(0)=
     |\overset{\circ}\psi\rangle\langle\overset{\circ}\psi|$
the Peres fidelity (\ref{Fidelity}) is simply the probability
\begin{equation}\label{Pure_in}
F_{\overset{\circ}\psi}(t)=
 |\langle \overset{\circ}\psi|{\hat f(t)}|\overset{\circ}\psi\rangle|^2=
 \left|\mathrm{Tr}\left[{\hat f(t)}\overset{\circ}\rho\right]\right|^2
\end{equation}
to survive in this state under influence of a chaotically evolving
perturbation ${\hat{\cal V}}(\tau)$ till the time $t$. When the
evolution starts from a mixed state
${\hat\rho}(0)=
    \sum_k  p_k
         |\overset{\circ}\psi_k\rangle \langle\overset{\circ}\psi_k|,\,\,
    \sum_k p_k=1$,
the expression (\ref{Pure_in}) can be generalized in two different
ways. The first of them leads to the standard definition
(\ref{Fidelity}) that can be rewritten identically in the form
\begin{equation}\label{Fidelity+}
   F(t)=\frac{1}{\mathrm{Tr}{\hat\rho^2(0)}}
            \sum_{k,k^\prime} p_{k} p_{k^\prime}W_{k k^\prime}(t)
\end{equation}
where the quantities
$W_{k k^\prime}(t)=
     |\langle \overset{\circ}\psi_k|{\hat f}(t)|
     \overset{\circ}\psi_{k^\prime}\rangle |^2$
are  probabilities of transitions induced by the unitary
transformation ${\hat f}(t)$.
The influence of coherent effects is hidden in the dynamics of the
complex matrix elements
$f_{k k'}(t)=\langle \overset{\circ}\psi_k|{\hat f}(t)|
                 \overset{\circ}\psi_{k^\prime}\rangle$.

Another way of generalization is suggested by the experimental
configuration with periodically kicked ion traps proposed in
\cite{Gardi97}. In such Ramsey type interferometry experiments
one directly accesses the fidelity amplitudes (see \cite{Gardi97}
rather than their square moduli. Motivated by this consideration,
we will consider further the quantity
\begin{equation}\label{Aledg}
{\cal F}(t)=
  \left|\mathrm{Tr}\left[{\hat f(t)}{\hat\rho}(0)\right]\right|^2=
  \Big|\sum_k p_k f_k(t)\Big|^2=
  \sum_k p^2_k F_k(t)+\sum_{k,k'}(1-\delta_{kk'}) p_k
  p_{k'}f_k(t)f_{k'}^*(t)\,,
\end{equation}
that is obtained by directly extending the formula (\ref{Pure_in}).
Below we refer to this new quantity as {\it allegiance}. The first
term in the r.h.s. is the sum of fidelities
$F_k=|f_k|^2=|\langle \overset{\circ}\psi_k|{\hat f}|
           \overset{\circ}\psi_k\rangle|^2$
of the individual pure initial states with weights $p_k^2$, while the
second, interference term, depends on the relative phases of fidelity
amplitudes. If the number $K$ of pure states
$|\overset{\circ}\psi_{k}\rangle$ that form the initial mixed state
is large, $K\gg 1$, so that $p_k\backsimeq 1/K$ for $k\leqslant K$
and zero otherwise, the first term is $\backsim 1/K$ at the initial
moment $t=0$ while the second term $\backsim 1$. Therefore, in
the case of a wide mixture, the decay of the function ${\cal F}(t)$ is
determined by the second sum of interfering contributions.
Therefore, in contrast to the standard Peres fidelity (\ref{Fidelity},
\ref{Fidelity+}), the allegiance ${\cal F}$ directly {\it accounts for
the quantum interference} and can be expected to retain quantal
features even in the deep semiclassical region. This is not, however,
the case as will be shown below.

Analytical calculation of the pure-state fidelity
$F_{\overset{\circ}\alpha}(t)$ for a pure coherent quantum state
$|\overset{\circ}\alpha\rangle$ as well as the allegiance
${\cal F}(t)$ for an incoherent mixed state can be performed with
the help of expressing the both quantities in terms of the Feynman's
path integral in the oscillator's phase plane. A method of
semiclassical evaluation of this integral has been worked out in
\cite{Sokol84}. Referring the reader to this paper for all technical
details we present below the main results of the calculations
\cite{Sokol07}.

These results are quite different it the two cases of our interest. If
the initial state is a pure coherent one
$|\overset{\circ}\alpha\rangle$ the fidelity amplitude is found to be
\begin{equation}\label{fidocoherent}
f_{\overset{\circ}\alpha}(t)=\frac{2}{\pi\hbar}\int d^2\delta\,
     e^{-\frac{2}{\hbar}|\delta|^2}\exp\left\{i\frac{\sigma}{2}
     \left[{\tilde\theta_c(t)}-
     {\overset{\circ}{\tilde\theta_c}}\right]\right\},
\end{equation}
where the phase
${\tilde\theta_c(t)}=\theta_c(\omega_0-2|\delta|^2;
  \overset{\circ}\alpha^*+\delta^*,
  \overset{\circ}\alpha+\delta;t)=\int_0^t
  d\tau[\omega_0-2|\delta|^2+2{\tilde I}_c(\tau)$.
It should be stressed that the fidelity
 $F_{\overset{\circ}\alpha}(t)=|f_{\overset{\circ}\alpha}(t)|^2$
does not decay in time at all if the quantum fluctuations described
by the integral over $\delta$ in (\ref{fidocoherent}) are neglected.

On the initial stage of the evolution
$t\ll\frac{1}{\Lambda}\ln\frac{2}{\varepsilon}$, while the phases
${\tilde\theta_c(t)}$ are not yet perfectly randomized the fidelity
$F_{\overset{\circ}\alpha}$
decays, because of classical exponential instability,
super-exponentially:
$$F_{\overset{\circ}\alpha}(t)\approx
\exp\left(-\frac{\varepsilon^2}{4\hbar}e^{\Lambda t}\right),$$
\cite{Silve02, Iomin04}. During this time the contribution of
the averaging over the initial Gaussian distribution in the
classical $\overset{\circ}\alpha$ phase plane dominates while the
influence of the quantum fluctuations of the linear frequency
remains negligible. Such a decay has, basically, a classical nature
\cite{Eckha03} and the Planck's constant appears only as the size of
the initial distribution. For larger times the quantum fluctuations
reduce the fidelity decay to exponential law
$F_{\overset{\circ}\alpha}(t)=\exp(-2\Lambda t).$

The situation changes dramatically if the initial state is an
incoherent mixture. More precisely, we consider a mixed initial state
represented by a Glauber's diagonal expansion
\cite{Glaub63,Glaub63a}
${\hat\rho}(0)=~\int d^2\overset{\circ}\alpha {\cal
P}(|\overset{\circ}\alpha-
\overset{\circ}\alpha_c|^2)|\overset{\circ}\alpha\rangle
\langle\overset{\circ}\alpha|$ with a wide positive definite weight
function ${\cal P}$ which covers a large number of quantum cells.
Note that here and in the following we assume that the initial
mixture is isotropically distributed in the phase plane around a fixed
point $\overset{\circ}\alpha_c\,$,
with the density
${\cal P}_{\overset{\circ}\alpha_c}(\overset{\circ}\alpha^*,
\overset{\circ}\alpha)={\cal P}(|\overset{\circ}\alpha-
\overset{\circ}\alpha_c|^2)$. Then allegiance equals \cite{Sokol07}
${\cal F}(t;{\overset{\circ}\alpha_c})=
|f(t;{\overset{\circ}\alpha_c})|^2$,
where
\begin{equation}\label{ampmx}
\begin{array}{c}
  f(t; {\overset{\circ}\alpha_c})=\int d^2\overset{\circ}\alpha\,
  {\cal P}(|\overset{\circ}\alpha-\overset{\circ}\alpha_c|^2)
  f_{\overset{\circ}\alpha}(t)\\
  \approx\int d^2\overset{\circ}\alpha\,
  {\cal P}(|\overset{\circ}\alpha-
  \overset{\circ}\alpha_c|^2)\exp\left\{i\frac{\varepsilon}{2\hbar}
  \left[\theta_c(t)-\overset{\circ}\theta_c(0)\right]\right\}.
\end{array}
\end{equation}
This formula directly relates the decay of a \textit{quantum}
quantity, the allegiance, to that of correlation function of the
\textit{classical} phases. No quantum features are present in the
r.h.s. of (\ref{ampmx}).

Summarizing,
the decay pattern of the allegiance ${\cal F}(t)$ depends on the
value of the parameter $\sigma=\varepsilon/\hbar$. In particular, for
$\sigma\ll 1$, we recover the well known Fermi Golden Rule (FGR)
regime. Indeed, in this case the cumulant expansion can be used,
$\ln f(t; {\overset{\circ}\alpha_c})=
 \sum_{\kappa=1}^{\infty} \frac{(i\sigma)^{\kappa}}{\kappa!}
 \chi_{\kappa}(t)\,.$
All the cumulants are real, hence, only the even ones are significant.
The lowest of them,
\begin{equation}\label{qum}
\begin{array}{c}
 \chi_2(t)=\int_0^t d\tau_1\int_0^t d\tau_2\langle \left[
 I_c(\tau_1)- \langle I_c(\tau_1)\rangle\right]
 \left[ I_c(\tau_2)-\langle I_c(\tau_2)
 \rangle\right]\rangle\equiv\int_0^t d\tau_1 \int_0^t
 d\tau_2 K_I(\tau_1,\tau_2)\;,
\end{array}
\end{equation}
is positive. Assuming that the classical autocorrelation function
decays exponentially,
$K_I(\tau_1,\tau_2)=
 \langle\left(\Delta I_c\right)^2\rangle
 \exp\left(-|\tau_1-\tau_2|/\tau_I\right)$
with some characteristic time $\tau_I$, we obtain
$\chi_2(t)=2\langle\left(\Delta I_c\right)^2\rangle\tau_It=2Kt$
for the times $t>\tau_I$ and arrive, finally, at the FGR decay law
${\cal F}(t; {\overset{\circ}\alpha}_c) =\exp(-2\sigma^2Kt)$
\cite{Cerru02,Jacqu01,Prose02}. Here
$K=\int_0^{\infty}d\tau K_I(\tau,0)=
        \langle\left(\Delta I_c\right)^2\rangle\tau_I\,.$

The significance of the higher connected correlators
$\chi_{\kappa\geq 4}(t)$ grows with the increase of the parameter
$\sigma$. When this parameter roughly exceeds one, the cumulant
expansion fails and the FGR approximation is no longer valid. In the
regime $\sigma\gtrsim 1$, the decay rate of the function ${\cal
F}(t; {\overset{\circ}\alpha}_c)=\big|f(t;
{\overset{\circ}\alpha_c})\big|^2$ ceases to depend on $\sigma$
\cite{Sagde88} and coincides with the decay rate $1/\tau_c$ of
the classical correlation function (\ref{ClasCorrDecay}),
\begin{equation}\label{Ldecay}
{\cal F}(t; {\overset{\circ}\alpha}_c)=\exp(-t/\tau_c)\,.
\end{equation}
This rate is intimately related to the local instability of the chaotic
classical motion though it is not necessarily given by the Lyapunov
exponent $\Lambda$ itself. Quantum interference {\it does not show
up} at all.

\subsection{Ballistic electron quantum transport in the presence of weakly disordered background.}
\label{sec:BallisticElTransport}
    Finally, we will discuss the decoherence phenomenon in the
electron transport through an open ballistic quantum dot as this
problem is seen from the point of view of the general resonance
scattering theory \cite{Sokol10}. Peculiarities of this transport
reflect the properties of eigenstates of the quantum billiards, whose
spectra are highly nontrivial in the classicaly chaotic regimes.

The loss of coherence that is the main problem of our concern is
attributed below to interaction with a weakly disordered many-body
environment (``walls''). So, the whole our system consists of an open
cavity and walls and is described by the following non-Hermitian
effective Hamiltonian
\begin{equation}
  {\cal \hat H}=
    \left(
    \begin{array}{cc}
       {\cal H}^{(s)} & V^{\dag}\\
        V        & H^{(e)}\\
    \end{array}%
    \right).
\end{equation}
The upper left block stands for the non-Hermitian effective
Hamiltonian  of the irregularly shaped cavity (dot) with two similar
leads supporting each $M/2$ equivalent channels.
\begin{equation}
  {\cal H}^{(s)} = H^{(s)}-\frac{i}{2}A A^{\dag}
\end{equation}
This non-Hermitian effective Hamiltonian describes a set of
$N^{(s)}$ electron doorway resonance states with complex
eigenenergies ${\cal E}_n=E_n-\frac{i}{2}\Gamma_n$ separated
by mean level spacing $D$.
The Hermitian matrix $H^{(e)}$ represents the environment with a
very dense discrete spectrum of $N^{(e)}$ ($>>> N^{(s)}$)  real
energy levels $\epsilon_{e}$ (mean level spacing
$\delta<<<D$). These states get excess to the continuum
only due to the coupling $V$ to the doorway states in the cavity.

We exploit further the single particle approximation in the
environmental sector: $H^{(e)}\Rightarrow H^{(e)}_{sp}$. The
mean level spacing of a quasi-electron $d\propto 1/N^{(e)}_{sp}$
is much greater that the many-body spacing $\delta$ but still much
smaller than the doorway spacing, $ \delta\ll d\ll D$. The
interaction $V$  with to irregular impurities is
described by a rectangular $N^{(e)}_{sp}\times N^{(s)}$
matrix with random matrix elements
\begin{equation}\label{av_V}
  \langle V_{\nu n}\rangle=0, \quad
  \langle  V_{\mu m}^* V_{\nu n}\rangle =
     \frac{1}{2} \Gamma_s \frac{d}{\pi}\delta_{\mu\nu}\delta_{m n}
\end{equation}
The second relation defines the spreading width
\begin{equation}\label{Gamma_s}
  \Gamma_s=2\pi\frac{\langle | V|^2\rangle}{d}
\end{equation}
that satisfies the condition ${\Gamma_s}\gg d$, so that the influence
of the disorder lies beyond validity of the standard perturbation
theory.

The unitary $M\times M$ scattering matrix has the form
\begin{equation}
   S(E)=I -i {\cal T}(E)=I-i {A^{\dag}}{\cal G}_D(E)A.
\end{equation}
Therefore the evolution of a scattered electron inside the cavity is
described by the $(N^{(s)}\times N^{(s)})$ doorway resolvent
(doorway propagator)
\begin{equation}\label{D_way_prop}
  {\cal G}_D(E)=\frac{ I}{E-{\cal H}^{(s)}-\Sigma(E)}.
\end{equation}
Here the Hermitian $N^{(s)}\times N^{(s)}$-matrix
\begin{equation}\label{Sigma}
  \Sigma(E)= V^{\dag}\frac{I}{E - H_{sp}^{(e)}}V=V^{\dag}G_{sp}^{(e)}(E)V
\end{equation}
accounts for transitions cavity $\leftrightarrow$ environment. Being
averaged over the random coupling amplitudes $V$, this matrix  is,
with accuracy $1/N^{(e)}_{sp}$, diagonal and is proportional to
the trace in the single-quasi-particle space.
\begin{equation}\label{Sigma_av}
  \Sigma(E)\Rightarrow\frac{1}{2}{\Gamma_s}\,g(E);\quad g(E)=\frac{d}{\pi}Tr G_{sp}^{(e)}(E).
\end{equation}
As a result a given doorway resonance is fragmented in a large
number of narrow resonances whose complex energies are found by
solving the equation
\begin{equation}\label{Fine_str}
  {\cal E}_{\nu}^n -{\cal E}_n-\frac{1}{2}\Gamma_s
      g({\cal E}_{\nu}^n)=0,
\end{equation}
so that $\Gamma_s/d$ fine structure resonances originate from any
given doorway doorway state.

The transition matrix transforms now to
\begin{equation}\label{T_exact}
  {\cal T}^{ab}(E)=\sum_n
    \frac{{\cal A}_n^a {\cal A}_n^b}{E-{\cal E}_n
     -\frac{1}{2}{\Gamma_s} g(E)}
  =\sum_{\nu}
    \frac{\tilde{\cal A}_{\nu}^a \tilde{\cal A}_{\nu}^b}{E-
    {\cal E}_{\nu}}.
\end{equation}
The resulting transition amplitudes are now sums of interfering
contributions of all narrow fine-structure resonances. The new pole
residues are complex and, therefore, interfere! \textit{No loss of
coherence on this stage!}\\

\subsubsection{Time delay.}
\label{sec:Timedelay}
    The resonant Smith time delay matrix $Q=-iS^{\dag}dS/dE$ can be expressed
\cite{Sokol97} in terms of the vectors $b^{(s,e)}$ of the intrinsic part of
the total scattering wave function. A straightforward calculation gives $Q\!=\!{b^{(s)}}^{\dag}b^{(s)}+{b^{(e)}}^{\dag}b^{(e)} \!=\!Q^{(s)} + Q^{(e)}$ where the vectors
\begin{equation}\label{bb}
\quad b^{(s)}(E)={\cal G}^{(s)}_{\textsc{d}}(E)A^{(s)}\,;
\quad b^{(e)}(E)=G_{sp}^{(e)}(E)Vb^{(s)}(E)
\end{equation}
have dimensions $N^{(s)}\times M$ and $N_{sp}^{(e)}\times M$
respectively. The two contributions $Q^{(s,e)}$ correspond to the modified due to the
interaction with the background time delay within the dot and delay because of the
virtual transitions into the background. In particular, the diagonal elements of the
resonant Smith matrix give the norms of the internal parts of the scattering wave
function initiated in specific channels. Finally, the typical scattering duration called
the Wigner delay time is represented by the following weighted-mean quantity
\begin{equation}\label{Wdt}
\tau_W(E)=\frac{1}{M}{\rm Tr}\, Q(E)\,.
\end{equation}

After averaging over the interaction $V$ we arrive in the main approximation to
\begin{equation}\label{Lam}
Q(E)=\Lambda(E)\,{b^{(s)}}^{\dag}b^{(s)}=\Lambda(E)
Q^{(s)}(E)\,;\quad \Lambda(E)=1+\frac{1}{2}\Gamma_s
l^{(e)}\;(E)
\end{equation}
where $l(E)$=$\frac{d}{\pi}$$\rm Tr$ $\left[{G_{sp}^{(e)}}^{\dag}(E) G_{sp}^{(e)}(E)\right]$.

In fact, however, the spectrum of the fine-structure resonances is
extremely dense so that this structure cannot be resolved experimentally. Only
quantities averaged over some energy interval $d\ll\Delta E\ll D$ are observed.
\begin{equation}\label{Csecn, Tdelay_av}
\begin{array}{c}
\overline{\sigma^{ab}(E)}=
    \frac{1}{\Delta E}\int_{E-\frac{1}{2}\Delta E}^{E+\frac{1}{2}\Delta E}
    dE'\,\big|{\cal T}^{ab}(E')\big|^2\,;\\
\overline{\tau_W(E)}=
    \frac{1}{\Delta E}\int_{E-\frac{1}{2}\Delta E}^{E+\frac{1}{2}\Delta E}
    dE'\,\tau_W(E')\,.
\end{array}
\end{equation}

Neglecting unobservable spectral fluctuations of the background we assume a rigid
spectrum with equidistant levels $\epsilon_{\mu}\!=\!\varepsilon_0\!+\!\mu d\!$
(the {\it picket fence} approximation). One of the advantages of this uniform model
is that the loop functions $g(E), l(E)$ can be calculated explicitly \cite{Sokol97}
\begin{equation}\label{g,l}
g(E)=\cot\left(\frac{\pi E}{d}\right)\,,
\quad l(E)=\frac{\pi}{d}\sin^{-2}\left(\frac{\pi E}{d}\right)\,.
\end{equation}

\subsubsection{Isolated doorway resonance near the Fermi energy.}
\label{sec:IsolatedDoorwayRes}
    If the incoming electron with the energy $E\approx E_{res}$ excites
an isolated ($\Gamma=\sum_c\Gamma^c\ll D$) resonance state with
the energy $E_{res}\approx 0$ very close to the Fermi surface in
the environment, the transition cross section turns out to be equal to
\begin{equation}\label{TrCrossSec}
\sigma^{ab}(E)=\Big|{\cal T}^{ab}(E)\Big|^2=
  \frac{\Gamma^a\Gamma^b}{\left[E-
  \frac{1}{2}\Gamma_s\cot\left(\frac{\pi E}{d}\right)\right]^2+
  \frac{1}{4}\Gamma^2}\,,
\end{equation}
when the Wigner time delay looks in this case as
\begin{equation}\label{Q1}
\tau_W(E)=\Gamma\,\frac{1+\frac{\pi\Gamma_s}{2d}\sin^{-2}\left(\frac{\pi
E}{d}\right)} {\left[E- \frac{1}{2}\Gamma_s\cot\left(\frac{\pi E}{d}\right)\right]^2+ \frac{1}{4}\Gamma^2}\,.
\end{equation}
The both quantities reveal strong fine-structure fluctuations on the typical scale
of the background level spacing $d$.

The fine-scale energy averaging yields then
\begin{equation}\label{AvTrCrossSec}
\overline{\sigma^{ab}(E)}=\frac{\Gamma^a\Gamma^b}{E^2+
  \frac{1}{4}\left(\Gamma+\Gamma_s\right)^2}+
  \frac{\Gamma^a\Gamma^b}{\Gamma}\frac{\Gamma_s}{E^2+
  \frac{1}{4}\left(\Gamma+\Gamma_s\right)^2}=\left(1+\frac{\Gamma_s}{\Gamma}\right)
  \frac{\Gamma^a\Gamma^b}{E^2+\frac{1}{4}\left(\Gamma+\Gamma_s\right)^2}\,
  \,.
\end{equation}
The averaging destroyed the coherence and decomposed the cross section into a sum
of two incoherent contributions. The first of them corresponds to excitation and
subsequent decay of the doorway resonance (widened because of leaking into the
environment) through one of the $M$ outer channels. The leakage effect is described by
additional shift in the upper part of the complex energy plane by the
distance $\frac{1}{2}\Gamma_s$. The second term accounts for the
particles re-injected back in the cavity from the background. There is no net
loss of the electrons. All of them escape finally via outer channels.

The environment looks from outside as a black box which swallows a particle and
spits it back in the cavity after some time. This time is characterized by the mean
Wigner time delay that also consists of two contributions,
\begin{equation}\label{av_t_delay}
\overline{\tau_W}(E)=\frac{\Gamma+\Gamma_s}{E^2
  +\frac{1}{4}(\Gamma+\Gamma_s)^2}+\frac{2\pi}{d}\;.
\end{equation}
The first term describes the delay on the damped by the internal ``friction''
resonance level inside the dot when the second one accounts for the electrons
delayed in the environment by the time $\tau_d=\frac{2\pi}{d}$ proportional to
the quasi-particle's level density.

The conductivity of the device is proportional to the transport cross section
\begin{eqnarray}\label{Transp1}
  G(E)&=&\sum_{a\in 1,b\in 2}\overline{\sigma^{a b}(E)}=
    \frac{\Gamma_1\,\Gamma_2}{\Lambda(E)}+
    \frac{\Gamma_ 1\,\Gamma_2}{\Gamma_1+\Gamma_2}\,
    \frac{\Gamma_s}{\Lambda(E)}\\
  &=&T_{1 2}+\frac{T_{1 s}\,T_{s 2}}{T_{1 s}+T_{s 2}}
\end{eqnarray}
where $\Gamma_k=\sum_{c\in k}\Gamma^c$, $k=1,2$;
$\Gamma_1+\Gamma_2=\Gamma$ and
\begin{equation}\label{SubTrProb1}
T_{s k}(E)=\frac{\Gamma_s\,\Gamma_k}{\Lambda(E)}\,,\quad\,\,\,\,
\Lambda(E)=E^2+\frac{1}{4}\left(\Gamma+\Gamma_s\right)^2.
\end{equation}

The term $T_{1 2}$ describes transition from the first to the second
lead via the broadened intermediate doorway resonance when the
additional contribution incorporates the interchanges with the
environment. The latter can be naturally interpreted by introducing
an additional fictitious $(M+1)th$ channel that connects the
resonance state with the environment. The corresponding extended
scattering matrix remains unitary.

The found expression is formally identical to that obtained within the
framework of the B\"{u}ttiker's voltage-probe model \cite{Buetti86,Buetti88}
of the decoherence phenomenon. The corresponding dimensionless decoherence
rate equals in our case to $\gamma_s=\frac{2\pi}{D}\Gamma_s=\Gamma_s\tau_D$.

\subsubsection{Many-body effects.}
\label{sec:ManyBodyeff}
    The single-particle approximation used up to now is well justified
only when the scattering energy $E$ is very close to the Fermi
surface in the environment. For higher scattering energies,
many-body effects should be taken into account. They show up, in
particular, in a finite lifetime of the quasi-electron with the energy
$E>E_F=0$. The simplest way to account for this effect is to
attribute some imaginary part to the quasi-particle's energy,
$\varepsilon_{\mu}=\mu d - \frac{i}{2}\Gamma_e$. The resonant
denominator looks then as \cite{Sokol97}
\begin{equation}\label{AbsProp}
{\cal D}_{res}(E)=E-E_{res}
-\frac{1}{2}\Gamma_s(1-\xi^2)\frac{\eta}{1+\xi^2\eta^2}+
\frac{i}{2}\left(\Gamma+\Gamma_s\xi\frac{1+\eta^2}{1+
\xi^2\eta^2}\right)
\end{equation}
where $E_{res}$ is the position of the doorway resonance and
the following notations
$$\xi=\tanh\left(\frac{\pi\Gamma_e}{2 d}\right),
\quad \eta=\cot\left(\frac{\pi E}{d}\right)$$
has been used.
The transport cross section $G(E)$ retains still its
form (\ref{Transp1}) but the subsidiary transition
probabilities looks now as
\begin{equation}\label{SubTrProb1Abs}
T_{sk}(E)\Rightarrow T_{sk}(E;\kappa)=
\frac{\Gamma_s\,\Gamma_k}{\Lambda(E;\kappa)}
\end{equation}
instead of (\ref{SubTrProb1}). The factor
\begin{equation}\label{eq4}
\frac{1}{\Lambda(E;\kappa)}=\frac{1}{\Lambda(E)}\,\frac{1}{1+
\kappa\frac{\Lambda(E)}{\Gamma\Gamma_s}},\quad \Lambda(E)=(E-E_{res})^2+\frac{1}{4}(\Gamma+\Gamma_s)^2
\end{equation}
depends on the new parameter $\kappa$ that accounts for
inelastic effects in the background,
\begin{equation}\label{kappa}
\begin{array}{c}
\kappa=\frac{4\xi}{(1-\xi)^2}=e^{\gamma_e}-1\approx\\
\left\{\begin{array}{l}
\gamma_e\ll 1,\,\,\,\,\,\,\,\textrm{if}\,\,\,\,\,\,\,\,\,\,{\tau}_e\gg\tau_d\,,\\
e^{\gamma_e}\gg 1,\,\,\,\,\, \textrm{if}\,\,\,\,\,\,\,\,\,\,\tau_e<\tau_d\,,\\
\end{array}\right.
\quad(\gamma_e=\tau_d\Gamma_e)\,
\end{array}
\end{equation}
where ${\tau}_e=1/\Gamma_e$ is the lifetime of the quasi-electron
in the environment.

Strictly speaking, the assumed quasi-electron decay, that implies
infinite density of the final states in the background, seems to
destroys the unitarity of the scattering matrix in contradiction with
what has been stated before. In fact, a single-particle state once
excited in the environment with a very dense but, nevertheless,
discrete spectrum evolves after that quite similar to a quasi-stationary
state only till the time $2\pi/\delta\gg \tau_e=1/\Gamma_e$.
After this time, recovery of the initial non-stationary state begins.
An electron that carry in particular electric charge preserves to a certain
extent its individuality in the environment. It can lose, because of the
many-body effects, only a part of its energy but not the charge and inevitably
returns sooner or later in the cavity and  escapes finally via one of the
outer channels. There exists a good probability for the electron to be re-emitted
in the cavity with some intermediate energy $E_{out}<E_{in}\approx E_{res}$ within
the much shorter time interval $\tau_d=\frac{2\pi}{d}$. The portion of energy lost
by such a retarded electron dissipates inside the environment. As a result, the
background temperature jumps slightly up during each act of the scattering. However,
supposing that the environment system is bulky enough, we can disregard this
very slow increase of the environment temperature. Alternatively, we can suppose
that a special cooling technique is in use.

Near the doorway resonance energy $E_{res}$ the influence of the
finite lifetime effects is negligible within the range
$0\leqslant\kappa\lesssim\kappa_c=
\frac{4\Gamma\Gamma_s}{(\Gamma+\Gamma_s)^2}$. The critical
value $\kappa_c$ reaches it's maximum possible, $\kappa_c=1$,
when $\Gamma=\Gamma_s$ and becomes small if one out of the two
widths noticeably exceeds another. In these cases the interval of
weak absorption is very restricted and the absorption begins to play
an important role. If the resonance is so narrow that
$\Gamma\ll\Gamma_s$, then $\kappa_c\approx
4\frac{\Gamma}{\Gamma_s}\ll 1$ and the subsidiary probabilities
(\ref{SubTrProb1Abs}) at the resonance energy $E=E_{res}$
and $\kappa\gtrsim\kappa_c$ are small,
$T_{sk}(E=E_{res};\kappa)\approx\frac{16}{\kappa}
 \frac{\Gamma\Gamma_k}{\Gamma_s^2}\lesssim
 \frac{16}{\kappa_c}
 \frac{\Gamma\Gamma_k}{\Gamma_s^2}\approx
 4\frac{\Gamma_k}{\Gamma_s}\ll 1$.
On the other hand, the very quasi-particle concept is self-consistent
only if $\gamma_e=\tau_d\Gamma_e\lesssim 1$ so that the
physically feasible interval of the strong absorption regime is
$\kappa_c\approx
 4\frac{\Gamma}{\Gamma_s}\lesssim\kappa\lesssim 1$.
In this interval, the subsidiary transition probabilities (\ref{SubTrProb1Abs})
are much smaller than the direct transport probability $T_{1 2}$ thus signifying
suppression of quantum coherence, the degree of suppression being quantitatively
characterized by the decoherence rate $\gamma_s\equiv\frac{2\pi}{D}\Gamma_s$.
Another possible method of accounting for the absorbtion has been proposed
in \cite{Efeto95}. The absorption is modelled in this case by including in
the Hamiltonian a spatially uniform imaginary potential with the strength $-\frac{i}{2}\gamma_{\phi}$. Results of this two approaches become equivalent
with the obvious identification $\gamma_{\phi}=\gamma_s$.

\subsubsection{Overlapping doorway resonances.}
\label{sec:OverlapDoorwayRes}
    A number of overlapping doorway states can be excited if the
incoming electron energy $E_{in}$ appreciably exceeds the Fermi
energy. As before, the cross sections averaged over the fine
structure scale consist of incoherent contributions of directly
scattered and penetrated into the environment and then re-emitted
particles.
\begin{equation}
  \overline{\sigma^{{ab}}(E)}=
    \sigma_d^{ab}(E)+\sigma_r^{ab}(E),
\end{equation}
where the direct and re-emitted contributions are
\begin{eqnarray}
  \sigma_d^{ab}(E)&=&
    \Big|\sum_n \frac{{\cal A}_n^a {\cal A}_n^b}{{\cal D}_n(E)}\Big|^2,\\
  \sigma_r^{ab}(E)&=&\Gamma_s \int_0^\infty dt_r\,\sigma_r^{ab}(E;t_r),\\
   &&\sigma_r^{ab}(E;t_r)=
     \Big|\sum_n\frac{{\cal A}_n^a {\cal A}_n^b}{{\cal D}_n(E)}\,e^{-i{\cal E}_n t_r}\Big|^2,\\
   &&{{\cal D}_n(E)}=E-E_n+\frac{i}{2}\left(\Gamma_n+\Gamma_s\right).
\end{eqnarray}
The particles delayed within environment for different times
contribute incoherently.

Since the electron motion in the cavity is supposed to be
classically chaotic the ensemble averaging $\langle ...\rangle$
in the doorway sector is appropriate. It is easy to see that, as
long as the inelastic effects in the background are fully neglected,
such an averaging perfectly eliminates dependence of all mean cross
sections on the spreading width. Indeed, the ensemble averaged
cross section is expressed in terms of the S-matrix two-point
correlation function
$C_V^{a b}(\varepsilon)=C_0^{a b}(\varepsilon-i\Gamma_s)$ as
\begin{equation}\label{AnsEvCrossSec_d}
\langle\sigma_d^{ab}(E)\rangle=
  C_V^{a b}(0)=C_0^{a b}(-i\Gamma_s)=
  \int_0^{\infty} dt\,e^{-\Gamma_s t}\,K_0^{a b}(t)\,.
\end{equation}
The subscript $V$ indicates the coupling to the background and
the function $K_0^{a b}(t)$ is the Fourier transform of the
correlation function $C_0^{a b}(\varepsilon)$. On the other
hand, it is easy to show that
\begin{equation}\label{AnsEvCrossSec_r}
  \langle\sigma_r^{ab}(E;t_r)\rangle=\int_0^{\infty} dt\,
  e^{-\Gamma_s t}\,K_0^{a b}(t+t_r)\,.
\end{equation}
Therefore, finally,
\begin{equation}\label{AnsEvCrossSec}
\begin{array}{c}
  \langle\overline{\sigma^{ab}(E)}\rangle=
  \int_0^{\infty} dt\,e^{-\Gamma_s t}\,K_0^{a b}(t)\\+
  \Gamma_s\int_0^{\infty} dt_r\int_0^{\infty} dt\,e^{-\Gamma_s t}\,
  K_0^{a b}(t+t_r)=
\int_0^{\infty} dt\,\,K_0^{a b}(t)=\langle\sigma_0^{ab}(E)\rangle \,.
\end{array}
\end{equation}
The ensemble averaging, being in fact equivalent to the energy
averaging over the doorway scale $D$, suppresses all interference
effects save the elastic enhancement because of the time reversal
symmetry. The latter effect manifests itself in the weak localization phenomenon. The time reversal symmetry is violated owing to the energy absorption in the environment.

\subsubsection{Energy absorption and suppression of the weak localization.}
\label{sec:EnergyAbsorp&WeakLoc}
    Taking into account eq. (\ref{AnsEvCrossSec}) we rewrite the
ensemble-averaged cross sections as
$\langle\overline{\sigma^{ab}(E)}\rangle=
  \langle\sigma_0^{ab}(E)\rangle+\Delta\sigma^{ab}(E;\kappa)$
where the second contribution can be reduced \cite{Sokol10}
to the following compact expression
\begin{equation}\label{Delta_sigma}
\Delta\sigma^{ab}(E;\kappa)=
-\sqrt{\frac{\kappa \Gamma_s}{4}}\int_0^{\infty}
\frac{dt}{\sqrt{-\frac{d}{dt}+\frac{\kappa}{4\Gamma_s}
\left(\frac{d}{dt}-\Gamma_s\right)^2}}\,K_0^{a b}(t)\,.
\end{equation}
Being presented in such a form, this result is equally valid for both
the orthogonal (GOE, time reversal symmetry) as well as the unitary
(GUE, no time reversal symmetry) cases.

To simplify further calculation we will consider the case of an appreciably
large number $M\gg 1$ of statistically equivalent scattering channels, all
of them with the maximal transmission coefficient $T=1$. Then, first, the
channel indices $a, b$ can be dropped. And, second, the characteristic decay
time $t_W=1/\Gamma_W=\tau_D/M$ (this time is called the {\it dwell time} when
the inverse quantity is known as the {\it Weisskopf width}) of the function
$K_0(t)$ is much shorter than the mean delay time $\tau_D=2\pi/D$.
Independently of time reversal symmetry, the function $K_0(t)$ is
real, positive definite, monotonously decreases with time $t$ and
satisfies the conditions $K_0(t<0)=0,\,\,K_0(0)=1$. This allows us to
represent this function in the form of the mean-weighted decay
exponent \cite{Sokol08b}:
\begin{equation}\label{LaplRepr}
K_0(t)=\int_0^{\infty}d\Gamma\,
    e^{-\Gamma t}\,w(\Gamma),\quad
     \int_0^{\infty}d\Gamma\,w(\Gamma)=
    K_0(0)=1\,.
\end{equation}
Rigorously speaking, the weight functions $w(\Gamma)$ have
different forms before ($t<\tau_D$) and after ($t>\tau_D$)
the Heisenberg time $\tau_D$. However contribution of the latter
interval is as small as $e^{-M}$ \cite{Sokol08b}.
Neglecting such a contribution we obtain in any inelastic channel
\begin{equation}\label{Final_2}
\Delta\sigma(E;\kappa)=
-\sqrt{\frac{\kappa \Gamma_s}{4}}
\int_0^{\infty}
d\Gamma\,\frac{w(\Gamma)}{\Gamma}\frac{1}{\sqrt{\Gamma+
\frac{\kappa}{4\Gamma_s}(\Gamma+\Gamma_s)^2}}\,.
\end{equation}
In the strong absorption limit
$\kappa\gg
  \frac{4\Gamma_s\Gamma_W}{(\Gamma_s+\Gamma_W)^2}$
the parameter $\kappa$ disappears from the found expression and
the latter reduces to
\begin{equation}\label{InfAbsCor}
\Delta\sigma(E;\kappa)\Rightarrow -\Gamma_s\int_0^{\infty}
d\Gamma\,\frac{w(\Gamma)}{\Gamma\,(\Gamma+\Gamma_s)}
=-\Gamma_s\int_0^{\infty} dt_r\int_0^{\infty} dt\,
  e^{-\Gamma_s t}\,
  K_0^{a b}(t+t_r)\,.
\end{equation}
According to Eqs. (\ref{AnsEvCrossSec_d}, \ref{AnsEvCrossSec})
this brings us to the result
\begin{equation}\label{Efet's_Limit}
\langle\overline{\sigma(E)}\rangle=\int_0^{\infty} dt\,
e^{-\Gamma_s t}\,K_0(t)=
\int_0^{\infty}d\Gamma\frac{w(\Gamma)}{\Gamma+\Gamma_s}=
\langle\sigma_d(E)\rangle .
\end{equation}
The averaged cross section approaches in this limit the value
$1/\Gamma_s$ {\it independently of the symmetry class} when the
spreading width $\Gamma_s$ noticeably exceeds the typical widths
contributing to the integral over $\Gamma$.

\begin{figure}
\begin{center}
\includegraphics[width=80mm,
   keepaspectratio=true]{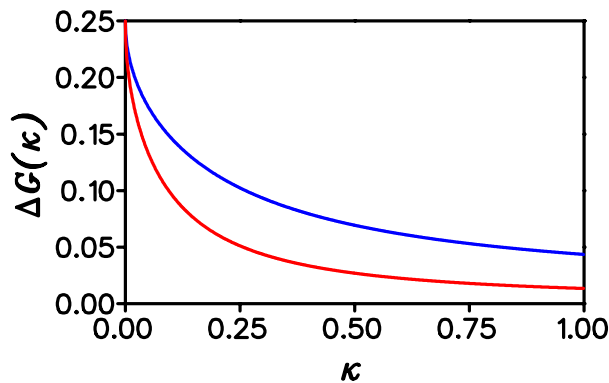}
\caption{Weak localization versus absorption parameter $\kappa$.
Lines correspond to $\gamma_s=25$ (blue) and
 $\gamma_s=64$ (red); $M=4$.}
\label{fig:WeakLoc}
\end{center}
\end{figure}

For the case of time reversal symmetry (GOE) the asymptotic
expansion \cite{Verba85} of the two-point correlation function gives
\cite{Sokol07}
\begin{equation}\label{W_TRS}
w^{(GOE)}(\Gamma)=\delta(\Gamma-\Gamma_W)-
\frac{2}{t_H}\delta'(\Gamma-\Gamma_W)
+\frac{M}{2t_H^2}\delta''(\Gamma-\Gamma_W)+...
\end{equation}
whereas for the case of absence of such a symmetry (GUE) similar
expansion result in
\begin{equation}\label{W_no_TRS}
w^{(GUE)}(\Gamma)=\delta(\Gamma-\Gamma_W)+...\,.
\end{equation}
In both these cases contributions of the omitted terms are estimated
as $O(1/(M^{-7/2}))$. With such an accuracy, the formula
(\ref{Final_2}) yields for the weak localization the expression
\begin{equation}\label{weak_loc_gen}
\begin{array}{c}
\Delta G\equiv G^{(GUE)}-G^{(GOE)}\\
= M_1M_2\left(2\frac{d}{d\mu}+
\frac{\mu}{2}\frac{d^2}{d\mu^2}\right)
\left\{\frac{1}{\mu}\left[1-\frac{\sqrt{\frac{\kappa\gamma_s}{4}}}
{\sqrt{\mu+\frac{\kappa}{4\gamma_s}(\mu+\gamma_s)^2}}\right]\right\}\Big|_{\mu=M}\\
\end{array}
\end{equation}
which is valid for arbitrary values of the parameters $\kappa$,
$\gamma_s$ and $M$. The unfolded explicit expression is a bit too
lengthy. Visualization of this result is presented in
Fig.~\ref{fig:WeakLoc} for two different values of the
(dimensionless) spreading width $\gamma_s$ and $M_1=M_2=2;
M=4$. The reduction of the difference $\Delta G$ displays
suppression of the quantum coherence. Note that the effect
becomes more pronounced as the number of channels decreases.

\section{Acknowledgments}
    We are very much obliged to Giuliano Benenti, Giulio Casati, and Yaroslav Kharkov
with whom we had the advantage of cooperation over a period of years. V.V.S. is
especially grateful to Vladimir Zelevinsky for long lasting friendship and collaboration.
This work is supported by the Ministry of Education and Science of the
Russian Federation (contract 14.B37.21.8408). Also we greatly appreciate
countenance by the RAS Joint scientific program "Nonlinear dynamics
and Solitons".

\bibliographystyle{abbrv}
\bibliography{dec-pre}

\end{document}